\documentclass[onecolumn,12pt,draftclsnofoot,journal]{IEEEtran}
\usepackage{cite}
\usepackage[mathscr]{eucal}
\usepackage{amsmath,lipsum}
\usepackage{amsfonts}
\usepackage{algorithm}
\usepackage{algorithmic}
\usepackage{hyperref}
\usepackage{float}
\usepackage{subfig} 
\usepackage{graphicx}
\usepackage{stfloats}
\usepackage{makecell}
\usepackage{color}
\usepackage{svg}
\usepackage[justification=centering]{caption}
\usepackage{makecell}

\hypersetup{hypertex=true,
colorlinks=true,
linkcolor=blue,
anchorcolor=blue,
citecolor=blue}

\newcommand{\BA}{\boldsymbol{\alpha}}

\newcommand{\CH}{\mathbf{g}}
\newcommand{\QQ}{\mathbf{Q}}
\newcommand{\BPHI}{\boldsymbol{\varphi}}

\newtheorem{definition}{Definition}
\newtheorem{theorem}{Theorem}

\newtheorem{lemma}{Lemma}
\newtheorem{remark}{Comment}

\begin{document}
\title{Channel Estimation with Dynamic Metasurface Antennas via Model-Based Learning}
\author{Xiangyu Zhang,~\IEEEmembership{Student Member,~IEEE,}
        Haiyang Zhang,~\IEEEmembership{Member,~IEEE,}
        Luxi Yang,~\IEEEmembership{Senior Member,~IEEE,}
        and~Yonina C.Eldar,~\IEEEmembership{Fellow,~IEEE,}
        \thanks{
X. Zhang, and L. Yang are with the School of Information Science and Engineering, and the National Mobile Communications Research Laboratory, Southeast University, and also with the Pervasive Communications Center, Purple Mountain Laboratories, Nanjing, China (e-mail: xyzhang@seu.edu.cn;  lxyang@seu.edu.cn); H. Zhang is with the School of Communication and Information Engineering, Nanjing University of Posts and Telecommunications, Nanjing, China. (e-mail: haiyang.zhang@njupt.edu.cn); Yonina C. Eldar is with the Weizmann Institute of Science, Rehovot, Israel (e-mail: yonina@weizmann.ac.il).
% Nir Shlezinger is with Ben-Gurion University of the Negev, Beer Sheva , Israel (e-mail: nirshl@bgu.ac.il)
}
}

\maketitle
\begin{abstract}

Dynamic Metasurface Antenna (DMA) is a cutting-edge antenna technology offering scalable and sustainable solutions for large antenna arrays. 
The effectiveness of DMAs stems from their inherent configurable analog signal processing capabilities, which facilitate cost-limited implementations. However, when DMAs are used in multiple input multiple output (MIMO) communication systems, they pose challenges in channel estimation due to their analog compression. 
In this paper, we propose two model-based learning methods to overcome this challenge. Our approach starts by casting channel estimation as a compressed sensing problem. Here, the sensing matrix is formed using a random DMA weighting matrix combined with a spatial gridding dictionary. We then employ the learned iterative shrinkage and thresholding algorithm (LISTA) to recover the sparse channel parameters. LISTA unfolds the iterative shrinkage and thresholding algorithm into a neural network and trains the neural network into a highly efficient channel estimator fitting with the previous channel.    
As the sensing matrix is crucial to the accuracy of LISTA recovery, we introduce another data-aided method, LISTA-sensing matrix optimization (LISTA-SMO), to jointly optimize the sensing matrix. LISTA-SMO takes LISTA as a backbone and embeds the sensing matrix optimization layers in LISTA's neural network, allowing for the optimization of the sensing matrix along with the training of LISTA. Furthermore, we propose a self-supervised learning technique to tackle the difficulty of acquiring noise-free data. Our numerical results demonstrate that LISTA outperforms traditional sparse recovery methods regarding channel estimation accuracy and efficiency. Besides, LISTA-SMO achieves better channel accuracy than LISTA, demonstrating the effectiveness in optimizing the sensing matrix.

% regarding channel estimation accuracy, LISTA outperforms 1 dB than optimization algorithms, and LISTA-SMO outperforms 3 dB than LISTA. Also, self-supervised learning achieves similar accuracy to supervised learning.
%

\end{abstract}

% \begin{IEEEkeywords}
% Channel estimation, Dynamic Metasurface Array (DMA), Model-based deep earning, Learning Iterative Shrinkage Thresholding
% Algorithm (LISTA),
% \end{IEEEkeywords}

\IEEEpeerreviewmaketitle
\section{Introduction}
To meet the requirements of future sixth-generation (6G) networks, large-scale antenna arrays are regarded as one of the critical physical-layer technologies \cite{bjornson_massive_2019}. However, in reality, constructing an antenna array with hundreds or even thousands of elements presents several challenges, including high fabrication costs, increased power consumption, limited physical size and shape, and deployment restrictions. To address these challenges, Dynamic Metasurface Arrays (DMA) have been proposed for the implementation of large-scale Multiple Input Multiple Output (MIMO) antenna arrays \cite{shlezinger_dynamic_2021}. DMAs are surface configurations of metamaterial radiators excited via waveguides, where one can alter the characteristics of each radiator. As a result, signal processing techniques, like compression and analog combination, are inherently integrated into the antenna analog processing chain, significantly reducing cost and power consumption. Additionally, DMAs allow for the fabrication of large-scale planar arrays and the stacking of a large number of elements~\cite{yoo2018enhancing, shlezinger_dynamic_2019, wang_dynamic_2021}. 

In recent years, there has been growing interest in the uses of DMAs for massive MIMO communications~\cite{shlezinger_dynamic_2019, wang_dynamic_2021, 9738442,jiang2022hybrid,you2022energy}. 
DMAs realize a form of hybrid MIMO antenna, which can be tuned to approach the achievable sum-rate of fully digital MIMO in some settings~\cite{shlezinger_dynamic_2019, shlezinger_dynamic_2021} and improve energy efficiency~\cite{you2022energy}. It was also shown that its analog processing capability could be leveraged to mitigate distortion due to low-resolution quantization~\cite{wang_dynamic_2021}, generate focused beams in near-field communications~\cite{9738442}, and be jointly adapted with smart programmable environments~\cite{jiang2022hybrid}. These findings highlight the potential benefits of DMAs in realizing massive and holographic MIMO systems~\cite{huang2020holographic}.  

While the inherent analog processing capability of DMAs contributes to their scalability and cost efficiency, it induces a notable challenge in acquiring CSI~\cite{shlezinger_dynamic_2019, wang_dynamic_2021}. Precisely, DMA's analog processing forms a unique compression where signals from elements on the same waveguides are combined. Consequently, the digital backend only observes this compressed signal. and the digital backend can only observe the compressed signal. Hence, the entire array's channel must be estimated from this compressed output at the digital backend. This presents an undetermined problem that has to be solved accurately and rapidly, i.e., within a coherence duration. 

Such challenges of analog compression are also present, with different constraints, in conventional phase shifter-based hybrid analog-digital antenna arrays \cite{alkhateeb_channel_2014}. Several works address the channel estimation problem in hybrid antenna arrays by employing compressed sensing (CS) theory \cite{eldar2015sampling, venugopal_channel_2017, rodriguez-fernandez_frequency-domain_2017}. 
Such techniques represent the channel in a sparse form, which often faithfully describes millimeter wave (mmWave) MIMO channels, and then recover it using CS algorithms such as orthogonal matching pursuit (OMP)~\cite{venugopal_channel_2017}, sparse Bayesian learning \cite{mishra_sparse_2017}, and iterative soft thresholding algorithm (ISTA) \cite{7891613,tibshirani1996regression,hastie2009elements}.  

Conventional sparse recovery approaches, such as ISTA and OMP, require fixing a known sensing matrix and tend to involve lengthy iterative optimization. Emerging model-based deep learning methodologies~\cite{shlezinger2020model,shlezinger2023model}, and particularly {\em deep unfolding}~\cite{monga2021algorithm}, provide a framework for leveraging data to alleviate such drawbacks of iterative optimizers. Recent works have studied the application of deep unfolding for phase shifter-based hybrid MIMO systems in the context of beamforming design~\cite{lavi2023learn,nguyen2023deep}, as well as for realizing fixed-latency sparse channel estimation~\cite{jin_adaptive_2021, chen_offset_2021}. Nonetheless, these existing works do not extend to the analog signal processing capabilities and constraints of the form of hybrid MIMO realized by DMAs.

Besides the recovering algorithm, the sensing matrix is crucial to the reconstruction accuracy \cite{Donoho2197}. When applying the CS in the channel estimation, the sensing matrix is the multiplication of the array weighting matrix and the sparsifying dictionary. For hybrid antenna arrays, a common approach for the sensing matrix uses random matrices as the array weighting matrix \cite{baraniuk2008simple} paired with a spatial gridding sparsifying dictionary \cite{jin_adaptive_2021}. This method straightforwardly aligns with the constraints of CS, such as the Restricted Isometry Property (RIP). However, this sensing matrix may not be optimal for channel estimation with DMAs. Two primary concerns arise. Firstly, the unique analog compression of DMA induces that a random DMA weighting matrix may not support the RIP. Secondly, the spatial gridding sparsifying dictionary inevitably introduces sparse representation errors \cite{9246294, 8370683}, which lowers the channel estimation accuracy. Given these challenges, optimizing the sensing matrix becomes vital for effective DMA channel estimation.

% Such optimized sensing matrices were proposed in \cite{ding2018dictionary, xie2020dictionary} for channel estimation, where channel measurements were used to learn a sparsifying dictionary from channel measurements rather than relying on a fixed over-complete Fourier matrix. However, these approaches often optimize the sensing matrix with singular value decomposition online, which induces a heavy computational burden, notably limiting its applicability for channel estimation, which has to be carried out anew on each coherence duration.

In this work, we tackle the challenge of rapid channel estimation for multi-user DMA-aided MIMO communications via two model-based algorithms. In particular, by leveraging the spatial sparse feature of the channel, we formulate the channel estimation as a CS problem, where the goal is to estimate the channel from a limited number of observation pilot signals. To achieve this, we construct a sensing matrix using a randomly weighted DMA matrix combined with a spatial gridding sparsifying dictionary. We then employ the Learning ISTA (LISTA) method, which adapts the traditional ISTA into a trainable model, as detailed in \cite{Gregor_ISTA}. Recognizing the challenges raised by the sensing matrix, we introduce another model-based algorithm with LISTA as the backbone, termed LISTA-Sensing Matrix Optimization (LISTA-SMO). This neural network integrates a sensing matrix optimizing layer into LISTA, enabling the optimization of the sensing matrix during LISTA training.

Next, we introduce a self-supervised learning (SSL) approach to train  LISTA-SMO. Compared with the supervised learning method in LISTA, SSL does not require the noiseless channels as labels, significantly reducing the training complexity and making it easier to implement in real systems. 

Finally, we provide extensive empirical evaluations to show the superiority of our algorithms. Particularly, numerical results show that the channel estimation accuracy of our proposed model-based learning methods is significantly better than the conventional non-learning methods. Also, thanks to optimizing the sensing matrix, LISTA-SMO achieves higher channel estimation accuracy than LISTA.
 
The rest of this paper is structured as follows: Section \ref{sec:system_model} outlines the operation of the DMA and the communication framework, presenting channel estimation in the context of a compressed sensing (CS) problem. Section \ref{sec:algorthem} revisits the LISTA method and applies it to address the channel estimation challenge. Also, this section explains how the accuracy of estimation is impacted by the sensing matrix constructed with a randomly weighted DMA matrix and a spatial gridding sparsifying dictionary. Following this, we detail the formulation of the sensing matrix optimization problem. Subsequently, we introduce our proposed LISTA-SMO, designed to optimize the DMA weighting matrix and sparsifying dictionary, and we discuss the Self-Supervised Learning (SSL) approach. Section \ref{sec:result} showcases our numerical findings, and Section \ref{sec:conclusion} provides concludes the paper.

Throughout the paper, we use the following notation. The bold type $\mathbf{a}$, $\mathbf{A}$ denote a vector and matrix, respectively. We use $\mathbf{A}^{\rm{T}}$ and $\mathbf{A}^{\rm{H}}$ to denote the transpose and the conjugate transpose of $\mathbf{A}$. $[\mathbf{a}]_{i}$ represents the $i$-th  element of $\mathbf{a}$. $[\mathbf{A}]_{i,j}$ represents the $i$-th row and $j$-th column element of $\mathbf{A}$. $\otimes$ represents the Kronick product. We use $\| \cdot \|_0, \| \cdot \|_1, \| \cdot \|_2$ for the $\ell_{0}$-norm, $\ell_{1}$-norm and $\ell_{2}$-norm, respectively, while $\mathbb{E}\{\cdot\}$ is the expectation operator, and $\mathbf{A}={\rm{diag}}(a_i)$ is a diagonal matrix whose diagonal elements are $\{a_i\}$.

\section{System Model and Problem Formulation}
\label{sec:system_model}

In this section, we present the DMA-based multi-user MIMO communication system and formulate the channel estimation problem. As shown in Fig.~\ref{systemmodel}, we consider a multi-user MIMO system consisting of a DMA-based base station (BS) and $K$ single-antenna users. We study the uplink channel estimation scenario, where users transmit orthogonal pilots to the BS, and BS estimates the channel according to the received pilots. In the following, we introduce the DMA processing in Subsection \ref{subsection-DMAmodel}. Then, the channel model is presented in Subsection \ref{subsection-ChannelModel}. We formulate channel estimation as a CS problem in Subsection \ref{subsection-problem}.

\subsection{Dynamic Metasurface Antennas}
\label{subsection-DMAmodel}
\begin{figure*}
\centering
\includegraphics[width=1\textwidth]{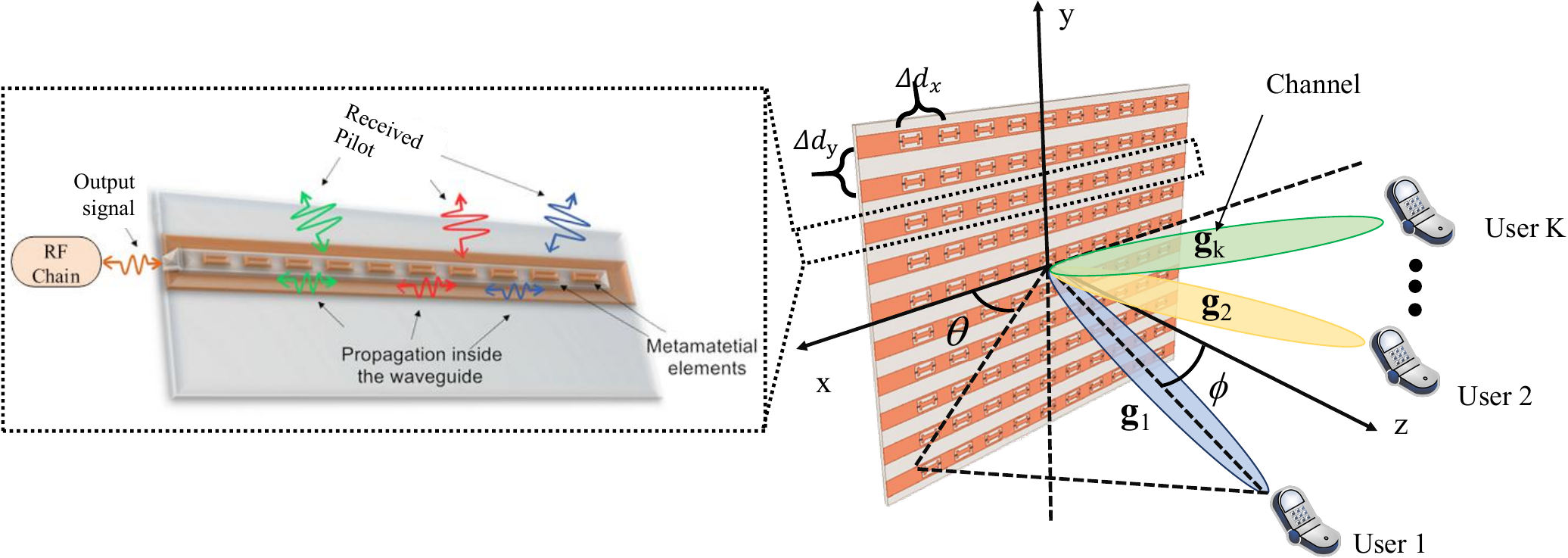}
\caption{The system model of multi-user channel estimation.}
\label{systemmodel}
\end{figure*}  

DMAs are two-dimensional configurations obtained by stacking one-dimensional waveguides with embedded metamaterial elements (meta-atoms)~\cite{shlezinger_dynamic_2019}. We consider a DMA consisting of $N_d$ microstrips, each with $N_e$ metamaterial elements (meta-atoms), for a total of $N = N_d\times N_e$ elements. DMAs receive signals via the metamaterial elements, whose amplitude and/or phase are adjustable. 

The signal at the output of the DMA, denoted by $\mathbf{z}_{\rm{RF}} \in \mathbb{C}^{N_d\times1}$, is given by  \cite{shlezinger_dynamic_2019, shlezinger_dynamic_2021, 9738442}
\begin{equation}
\label{DMA_transmitter}
\mathbf{z}_{\rm{RF}} = \mathbf{Q}  \mathbf{H} \mathbf{y},
\end{equation}
where $\mathbf{y} \in \mathbb{C}^{N\times1}$ is the signal observed by the $N$ elements, and $\mathbf{H} \in \mathbb{C}^{N\times N}$ is a diagonal matrix with diagonal elements  
\begin{equation}
 [\mathbf{H}]_{(n_d-1)N_e+n_e,(n_d-1)N_e+n_e} = {\rm exp}\left({-\rho_{n_d, n_e} (\alpha_{n_e}+j \beta_{n_e})}\right),
\end{equation}
which encapsulates the effect of signal propagation inside the microstrip and the propagation delay of $n_e$-th element in $n_d$-th microstrip. Here, $\alpha_n$ is the waveguide attenuation coefficient, $\beta_n$ is the wavenumber, and $\rho_{n, l}$ denotes the location. The matrix $\mathbf{Q} \in \mathbb{C}^{N_d \times N}$ represents the configurable weight of each metamaterial element, which can be written as
\begin{equation}
\label{Q_constrain}
[\mathbf{Q}]_{n_{1},\left(n_{2}-1\right) N_d+n_e}=\left\{\begin{array}{ll}
q_{n_{1}, n_e} & n_{1}=n_{2} \\
0 & n_{1} \neq n_{2}
\end{array} ,\right.\\
\end{equation}
where $q_{n, n_e} = \frac{j+{\rm{exp}}({j \varphi_{n, n_e}})}{2}$ is the weight on each element and $\varphi_{n, n_e}$ is the configurable phase \cite{2017Analysis}.

% following the Lorentzian constrain \cite{2017Analysis}, given by
% \begin{equation}
%     q \in \mathbb{Q} =\left\{\frac{j+e^{j \varphi}}{2} \mid \varphi \in[0,2 \pi]\right\}.
%     \label{Lorentzian-constrained}
% \end{equation}
% \eqref{Lorentzian-constrained} shows the DMA can adjust  to control the weight $q_{n, n_e}$.

As shown in Fig.~\ref{systemmodel}, we set the DMA in the XOY plane, where the microstrips are located in parallel with the x-axis. We use $\Delta d_x$ and $\Delta d_y$ to represent the space between the microstrip and the elements, while $\theta$ and $\phi$ denote the azimuth angle and elevation angle, respectively.

\subsection{Channel and Signal Model}
\label{subsection-ChannelModel}

In the context of uplink channel estimation, we assume that the $K$ users transmit mutually orthogonal pilot sequences to the BS. Thus, the channel estimation for each user can be considered to be independent. For the sake of simplicity and without any loss of generality, let's consider an arbitrary user. The channel $\mathbf{g}\in \mathbb{C}^{N\times 1}$ between an arbitrary user and DMA can be represented by \cite{el2014spatially,akdeniz2014millimeter}
\begin{equation}
\label{channel_model}
\mathbf{g} =
\sqrt{\frac{N}{L_p}}\sum_{l_p=1}^{L_p} a_{l_p} \mathbf{a}\left(\theta_{l_p}, \phi_{l_p}\right),
\end{equation}
where $L_p$ is the number of paths. We assume $L_p \ll N$ as the system is operated in the high-frequency band. $a_{l_p}$ is the channel gain of $l_p$-th path, and $\mathbf{a}\left(\theta_{l_p}, \phi_{l_p}\right) = \mathbf{a}_{x}\left(\theta_{l_p}, \phi_{l_p}\right)\otimes\mathbf{a}_{y}\left(\theta_{l_p}, \phi_{l_p}\right)$ represents the array response vector of $l_p$-th path, where $\mathbf{a}_{x}$ and $\mathbf{a}_{y}$ are formulated as 
\begin{equation}
\mathbf{a}_{x}\left(\theta_{l_p}, \phi_{l_p}\right)   = \frac{1}{\sqrt{N_e}}\left[1, e^{j d_x}, \ldots, e^{j d_x (N_e-1)}\right]^{\rm{T}},
\end{equation}
\begin{equation}
\mathbf{a}_{y}\left(\theta_{l_p}, \phi_{l_p}\right)  = \frac{1}{\sqrt{N_d}}\left[1, e^{j d_y}, \ldots, e^{j d_y (N_d-1)}\right]^{\rm{T}},
\end{equation}
with $d_x=2 \pi \frac{\Delta d_x}{\lambda}\cos\theta\sin\phi$ and $d_y= 2 \pi \frac{\Delta d_y}{\lambda} \sin \theta \sin \phi$ representing the propagation delay caused by the position difference of two elements and two microstrips, respectively. 

% For easy illustration, we assume users utilize the orthogonal pilot sequences, and the uplink channel for each user can be estimated independently. Then, 

Let $p$ be the uplink transmitted pilot, the received pilot observed by the DMA element $\mathbf{y}\in \mathbb{C}^{N\times 1}$ from the arbitrary user is 
\begin{equation}
\label{received_signal111}
\mathbf{y} =  \mathbf{g} p +  \mathbf{n},
\end{equation}
where $\mathbf{n}\in \mathbb{C}^{N\times 1}$ is noise, and each element of $\mathbf{n}$ follows the normal distribution, e.g.,  $[\mathbf{n}]_n \sim\mathcal{CN}( 0, \sigma^2)$. $\sigma$ is the variance of the normal distribution. 
Accordingly, the observed signal at the digital backend $\mathbf{z}\in \mathbb{C}^{N\times 1}$ is  
\begin{equation}
\label{received_signal2}
\mathbf{z} = \mathbf{Q} \mathbf{H}\mathbf{y} = \mathbf{Q} \mathbf{H}\mathbf{g} p +  \mathbf{n}_e,
\end{equation}
where $\mathbf{n}_e\in \mathbb{C}^{N\times 1}$ is equivalent noise, and each element of $\mathbf{n}_e$ follows the normal distribution, e.g.,  $[\mathbf{n}]_n \sim\mathcal{CN}( 0, \sum_{l = n_e}^{N_e}|q_{n,l}h_{n,l}|^2\sigma^2)$.

\subsection{Problem Formulation}
\label{subsection-problem}
% underdetermined problem since $N_d \ll N$. Nonetheless, the channel can be represented by the parameter set $[\alpha_{l_p}, \theta_{l_p}, \phi_{l_p}], l_p \in [1,..., L_p]$ as shown in \eqref{channel_model}. Therefore, it is more reasonable to estimate the parameter set $[\alpha_{l_p}, \theta_{l_p}, \phi_{l_p}]$ rather than $\mathbf{g}$. 

Our objective is to estimate $\mathbf{g}$ from $\mathbf{z}$, which is an underdetermined problem since $N_d \ll N$. One of the conventional ways to track this challenge is to exploit the channel's spatial sparse character and decompose channels into sparse representations, which is given by 
\begin{equation}
\mathbf{g} \simeq \mathbf{A}_G\boldsymbol{\alpha},
\end{equation}
where $\boldsymbol{\alpha} \in \mathbb{C}^{D\times 1}$ is sparse channel representation and $\mathbf{A}_G\in \mathbb{C}^{N\times D}$ is spatial partition sparse dictionary. The spatial partition sparse dictionary $\mathbf{A}_{\rm G}$ is constituted by the array response vectors on the space partition grid, given by $\mathbf{A}_{\rm G} = \left[
\begin{array}{ccc}
\mathbf{a}\left(\theta_{1}, \phi_{1}\right), &  \cdots,   &
\mathbf{a}\left(\theta_{D}, \phi_{D}\right)  
\end{array}\right]$, 
where $\theta_i$ and $\phi_i$ are the azimuth and elevation angles of the spatial partition; $D$ is the number of the space partition grid. From \eqref{channel_model}, we have the conclusion that $\|\boldsymbol{\alpha}\|_0 \simeq L_p$. Besides, we assume $D \gg L_p$, which implies $\boldsymbol{\alpha}$ is a highly sparse channel representation. 

Building on this sparse decomposition, we can estimate the channel $\mathbf{g}$ by recovering the sparse channel representation $\boldsymbol{\alpha}$ using compressed sensing techniques. Following  \cite{venugopal_channel_2017,mishra_sparse_2017,7891613}, we set the configurable phase of DMA as a random variable and formulate a randomly DMA weighting matrix $\mathbf{Q}_{\rm Ran}$. Then, the channel estimation problem can be formulated as a CS problem, which is given by
\begin{equation}
\begin{aligned}
\label{LASSO}
 ~  \arg\min_{\boldsymbol{\alpha}} ~& \|\boldsymbol{\alpha}\|_{0} \\
\text { s.t }~ & \|\mathbf{z} -  \mathbf{Q}_{\rm Ran} \mathbf{H}\mathbf{A}_G\boldsymbol{\alpha}  \|_2< \epsilon, \\
\end{aligned}
\end{equation}
where $\epsilon$ is the sparse constraint constant.  
% In \eqref{LASSO}, we use the symbol $\mathbf{A}$ to represent the sparsifying dictionary, which encompasses all types of such dictionaries. 
As the $l_0$ norm is non-convex, it is quite challenging to find the optimal solution to \eqref{LASSO}. By relaxing the $l_0$ norm to $l_1$ norm, we have a more tractable problem 
\begin{equation}
~ \min_{\boldsymbol \alpha}~{ \|\mathbf{z} -  \mathbf{Q}_{\rm Ran} \mathbf{H}\mathbf{A}_G\boldsymbol{\alpha} \|_2}  + \xi \|\boldsymbol{\alpha}\|_{1}
\label{eq_lasso},
\end{equation}
where $\xi$ is a regularization parameter.
Note that \eqref{LASSO} and \eqref{eq_lasso} share the same optimal solution when the configurable sensing matrix $\mathbf{\Psi}\left(\mathbf{Q},\mathbf{A}\right)$ satisfies the RIP \cite{eldar2012compressed}.

Problem \eqref{eq_lasso} can be solved with full-fledged iterative optimization methods, such as ISTA \cite{7510952,4959678}. However, these methods' performance heavily depends on the proper hyper-parameter, such as $\xi$. Besides, they have high computational complexity and converge slowly. To address these issues, we propose a model-based method to efficiently solve problem \eqref{eq_lasso} in the next section. 

% By giving the specific $\mathbf{Q}$ and $\mathbf{A}$
% , by giving a randomly DMA weighting matrix $\mathbf{Q}_{\rm Ran}$,

% $\mathbf{\Psi}\left(\mathbf{Q},\mathbf{A}\right) = \mathbf{Q} \mathbf{H}\mathbf{A}$ is the sensing matrix, which is the function of the DMA parameter matrix $\mathbf{Q}$ and the sparsifying dictionary $\mathbf{A}$

\section{DMA Channel Estimation via Model-Based Deep Learning}
\label{sec:algorthem}
In this section, we propose model-based deep learning methods to solve problem \eqref{eq_lasso}. Specifically, in Subsection \ref{sec:primary_LISTA}, we first employ the Learned ISTA (LISTA) to solve problem \eqref{eq_lasso}. Next, we state the sensing matrix that is compromised by $\mathbf{Q}_{\rm Ran}$ and $\mathbf{A}_G$ may hinder the accuracy of LISTA recovery. Hence, we formulate a sensing matrix optimization problem and propose a novel algorithm, called LISTA-sensing matrix optimization (LISTA-SMO), to solve the channel estimation and sensing matrix optimization simultaneously in Subsection \ref{subsection-LISTA-SMO}. Besides, considering the difficulty in obtaining the noise-less data, we propose the self-supervised learning method for our proposed methods in Subsection \ref{subsection-self-supervisedlearning}.

\subsection{LISTA with A Randomly Given Sensing Matrix}
\label{sec:primary_LISTA}
In this subsection, we employ the LISTA method to solve problem \eqref{eq_lasso} with a given sensing matrix as in \cite{venugopal_channel_2017,mishra_sparse_2017,7891613} that constructs the sensing matrix with randomly DMA weighting matrix $\mathbf{Q}_{\rm Ran}$ and spatial partition sparse dictionary $\mathbf{A}_{\rm G}\in \mathbb{C}^{N \times D}$. For notational simplicity, we use $\mathbf{\Psi}$ to represent the randomly given sensing matrix $\mathbf{\Psi}\left(\mathbf{Q}_{\rm Ran},\mathbf{A}_{\rm G}\right) $ in this subsection.
% By assuming the channel path is incident from these grids, the channel can be represented in a sparse form, given by 
% \begin{equation}
% \mathbf{g}\simeq\mathbf{A}_{G}\boldsymbol{\alpha}_{G},
% \end{equation}
% where $\boldsymbol{\alpha}_{G} = [\alpha_1, \cdots, \alpha_{D}]^{\rm{T}} \in \mathbb{C}^{D\times 1}$; $\alpha_d$ denotes the channel from $d$-th grid.

LISTA is a deep unfolding version of ISTA, which treats each iteration of ISTA as a neural network layer and learns the hyperparameter from data.  
Standard ISTA starts from an initial solution $\boldsymbol{\alpha}^{(0)}$ and iteratively performs proximal gradient descent
with respect to the objective function in \eqref{eq_lasso}. Specifically, the iteration can be formulated as \cite{beck2009fast} 
\begin{equation}
\label{ISTA_eq}
\BA{}^{(t+1)}=h_{(\eta)}\left(\left({\bf I}- \frac{1}{\lambda_{\max}} \mathbf{\Psi}^{{\rm{H}}}\mathbf{\Psi}\right)\BA{}^{(t)}+\frac{1}{\lambda_{\max}} \mathbf{\Psi}^{{\rm{H}}}\mathbf{z}\right),
\end{equation}
where
% \HYCmt{(x) should be written in this form $\left(x\right)$. } 
$\lambda_{\max}$ is the maximum eigenvalue of $\mathbf{\Psi}^{\rm{H}}\mathbf{\Psi}$; $h_{(\eta)}$ is the soft shrinkage function with threshold $\eta$, given by $[h_{(\eta)}]_g = \rm{sign}([\boldsymbol{\alpha}]_g)(|[\boldsymbol{\alpha}]_g|-\eta)_{+}$, where $[\cdot]_n$ is the $n$-th elements of vector; $\rm{sign}(\cdot)$ returns the sign of a scalar; $(\cdot)_{+}$ means $\max(\cdot,0)$; $\eta$ is a constant threshold.

% Equation (\ref{ISTA_eq}) shows that $\BA$ is hard to formulate as a closed-form expression because iterating time is unknown. 

\begin{figure*}[!h]
\centering
\includegraphics[width= 0.5\textwidth]{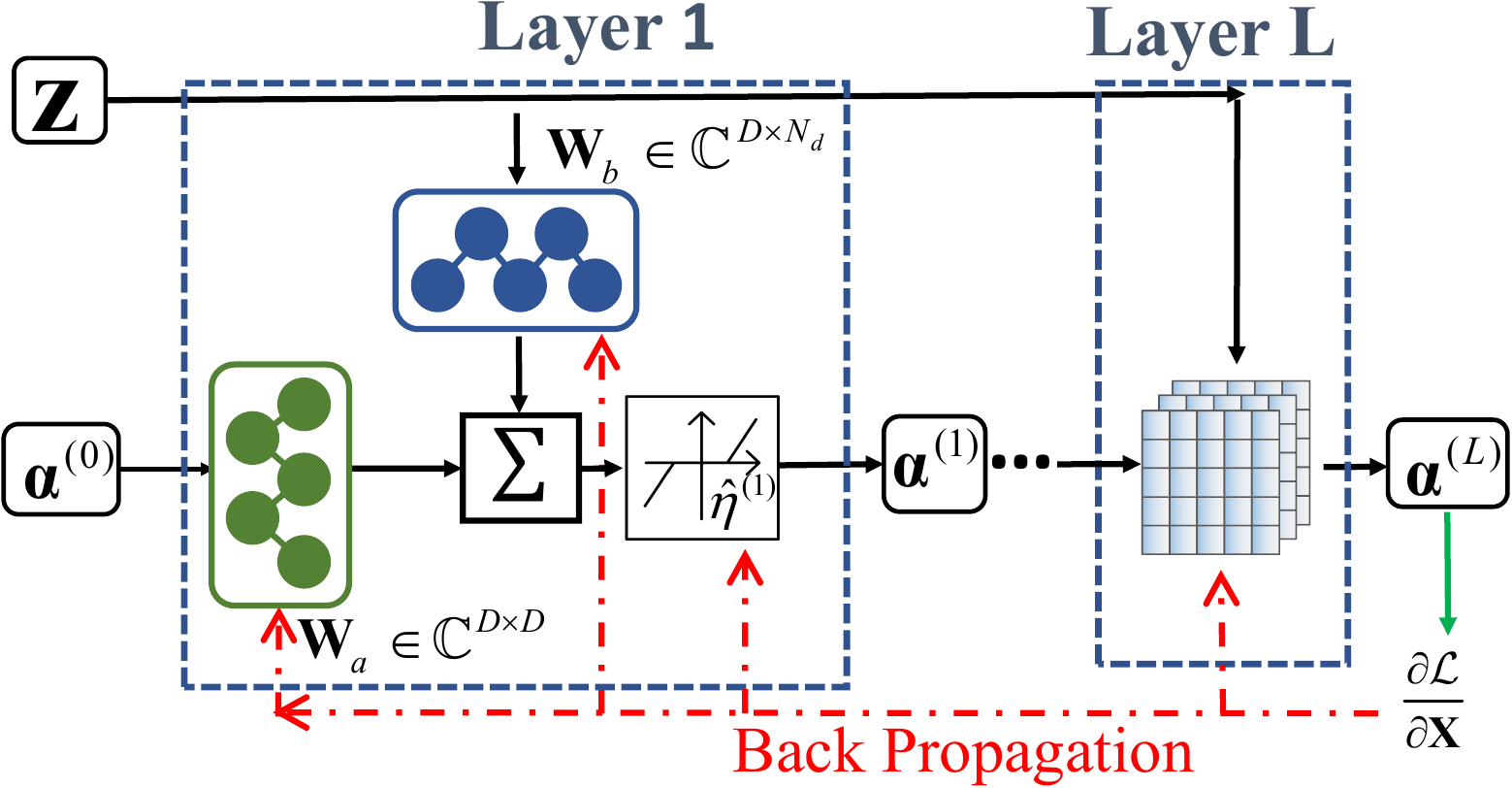}\\
\caption{The block diagram of LISTA.}
\label{fig-LISTA}
\end{figure*}

The recovering accuracy of ISTA heavily depends on the parameter of $\eta$ and the iterative time. Selecting the proper parameter for each $\BA$ is challenging. Besides, ISTA can also be computationally intensive, requiring hundreds or thousands of iterations to converge.
To address these limitations, we apply LISTA to solve problem \eqref{eq_lasso}, which learns the parameters such as threshold $\eta$ from the collected channel data set. Specifically, as shown in Fig.~\ref{fig-LISTA}, LISTA builds a $L$-layer feed-forward neural network with side connections, denoted as $\mathcal{N} ({\bf{z}}| \mathbf{W}_a, \mathbf{W}_b, \hat{\eta})$, whose layer can be formulated as \cite{Gregor_ISTA}   
\begin{equation}
\label{LISTAequation2}
\BA{}^{(l+1)}=h_{(\hat{\eta}^{(l)})}\left(\mathbf{W}_a\BA{}^{(l)}+\mathbf{W}_b\mathbf{z}\right),
\end{equation}
where $\mathbf{W}_a \in \mathbb{C}^{D \times D}$, $\mathbf{W}_b \in \mathbb{C}^{D \times N_d}$ and $\hat{\eta}^{(l)}$ are learnable parameter of the $l$-th layer corresponding to $({\bf I}- \frac{1}{\lambda_{\max}} \mathbf{\Psi}^{{\rm{H}}}\mathbf{\Psi})$, $\frac{1}{\lambda_{\max}} \mathbf{\Psi}^{{\rm{H}}}$ and $\eta$ in \eqref{ISTA_eq}, respectively. 

Fig.~\ref{fig-LISTA} also presents the forward-solve process and the backward-update process with the black and red lines, respectively. The forward process feeds the received pilot ${\bf{z}}$ to each layer as the weighted biased items. $\BA{}^{(l)}$ is the hidden layer feature flow through each layer. At the end layer, the neural network outputs the estimated sparse channel parameter $\boldsymbol{\alpha}^{(L)}$. 
The backward process updates the parameters of LISTA by minimizing the loss function 
$
    \mathcal{L} = \sum_{s=1}^S\| \boldsymbol{\alpha}^*_s - \boldsymbol{\alpha}^{(L)}_s \|_2,
$
on the LISTA's training data set, which is $\mathbb{D}_{\rm{LISTA}} = \left\{[\mathbf{z}_s, \boldsymbol{\alpha}^*_s]| \mathbf{z}_s = \mathbf{\Psi}\boldsymbol{\alpha}^*_s + \mathbf{n}, s = 1,..., S \right\}$. $\mathbf{z}_s$ and $\boldsymbol{\alpha}^*_s$ are the sample pair representing the received pilot and corresponding actual sparse-from channel. $\boldsymbol{\alpha}^{(L)}_s$ is the output corresponding to $\mathbf{z}_s$.

\begin{algorithm}[h]
\caption{DMA channel estimation with LISTA}
\label{alg2}
\begin{algorithmic}[1]
\STATE \textbf{Training Process}
\STATE \textbf{Initialize:} Initialize the parameter $\mathbf{W}_a$, $\mathbf{W}_b$ and $\hat{\eta}$.
\STATE \textbf{Input:} the data set $\mathbb{D}_{\rm{LISTA}} = \left[\{\mathbf{z}_s, \boldsymbol{\alpha}^*_s\}| \mathbf{z}_s = \mathbf{\Psi}\boldsymbol{\alpha}^*_s + \mathbf{n}, s = 1,...,S \right]$. 
\STATE \textbf{While} 
\STATE \quad Sample a batch of data from the data set $\mathbb{D}_{\rm{LISTA}}$;
\STATE \quad Calculate the output of neural network $\BA{}^{(l)}$ and calculate the loss $\mathcal{L}$;
\STATE \quad Update the parameter $\mathbf{W}_a$, $\mathbf{W}_b$ and $\hat{\eta}$. 
\STATE \textbf{Until} $\mathcal{L}$ converge. 
\STATE \textbf{Output:} The well-trained neural network $\mathcal{N}$.
\STATE \textbf{Inference Process}
\STATE Feed the received pilot $\mathbf{z}$ into the well-trained neural network $\mathcal{N}$ and obtain the estimated sparse channel $\boldsymbol{\alpha}^{(L)}$. 
\STATE The estimated channel $\mathbf{g}= \mathbf{A}_{\rm G}\boldsymbol{\alpha}^{(L)}$.
\end{algorithmic}
\end{algorithm}

After obtaining the estimated sparse channel parameter $\boldsymbol{\alpha}^{(L)}$ from the output at the end layer, the estimated wireless channel $\hat{\mathbf{g}}$ can be obtained from $\hat{\mathbf{g}} = \mathbf{A}_{\rm G}\boldsymbol{\alpha}^{(L)}$ directly. 
Therefore, the complete LISTA for solving problem \eqref{eq_lasso} and DMA channel estimation is summarized as Algorithm \ref{alg2}.

We note that LISTA is applicable to problem \eqref{eq_lasso}  only for the case of fixed sensing matrix, i.e., the DMA weighting matrix $\mathbf{Q}$ and the sparse dictionary $\mathbf{A}$ are given. Meanwhile, the performance of LISTA also heavily depends on the values of $\mathbf{Q}$ and $\mathbf{A}$. The randomly given sensing matrix may fail to satisfy the RIP condition, resulting in reduced estimation accuracy \cite{7096434}. To demonstrate the accuracy loss caused by the random weighting matrix $\QQ_{\rm Ran}$ and the spatial gridding sparse dictionary $\mathbf{A}_{\rm G}$, we derive the following two theorems.

\begin{theorem}
Assuming that each non-zero element of the random DMA weight matrix $\mathbf{Q}_{\rm Ran}$ have the random phase $\varphi \sim \mathcal{U}[0, 2\pi]$, then the sparse vector $\mathbf{\alpha}$ can be \textit{exactly} recovered from the noiseless pilot with a probability:
\begin{equation}
    p_{\rm rec} = \left[\frac{1}{2} + \frac{1}{2}\text{erf} \left( \frac{\sqrt{2} - N_e/2}{\sqrt{2}N_e/12} \right)\right]^{2L_p}, 
\end{equation}
where $\operatorname{erf}\left( z\right) =\frac{2}{\sqrt{\pi}} \int_0^z e^{-t^2} \mathrm{~d} t$ denotes the error function. 
\label{thm:unsuccessweight}
\end{theorem}
\begin{IEEEproof}
Please refer to Appendix \ref{AppendixA}.
\end{IEEEproof}

\begin{figure*}[!h]
\centering
\includegraphics[width=0.5\textwidth]{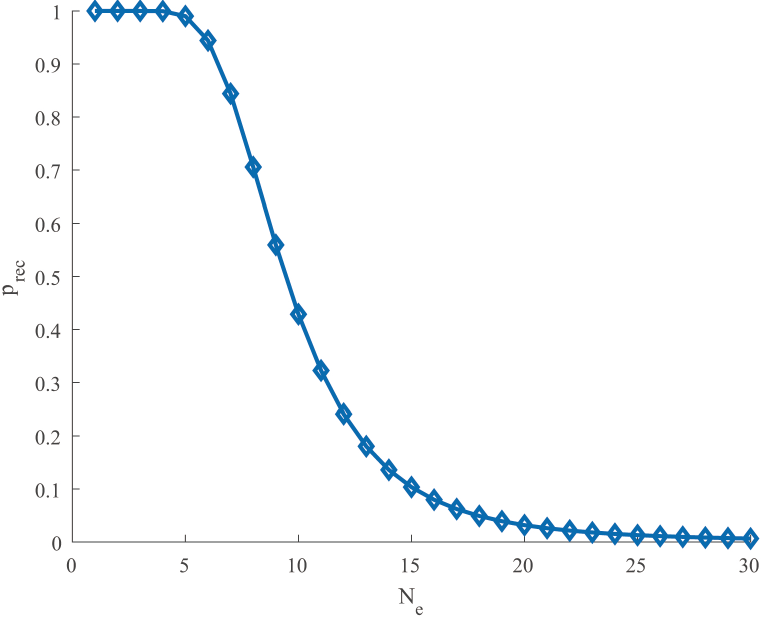}
\caption{The recovery probability $p_{\rm rec}$ versus $N_e$, under $N_d = 20$, $L_p=5$, and a randomly generated DMA weight matrix $\mathbf{Q}_{\rm Ran}$. }
\label{fig-recoverying_probability}
\end{figure*}

Theorem \ref{thm:unsuccessweight} implies that the recovery probability $p_{\rm rec}$ is a function of $N_e$ (the number of antenna elements on each microstrip). The larger the $N_e$ is, the lower the recovery probability $p_{\rm rec}$ is. To illustrate this point more clearly, we provide Fig.~\ref{fig-recoverying_probability}, which plots the recovery probability $p_{\rm rec}$  versus $N_e$ with a randomly generated DMA weight matrix $\mathbf{Q}_{\rm Ran}$.
 From Fig.~\ref{fig-recoverying_probability}, we can clearly see that when $N_e$ is small, the recovery probability $p_{\rm rec}$  approaches 1 even if the weight matrix $\mathbf{Q}$ is randomly generated. However, $p_{\rm rec}$  decreases rapidly when $N_e$ is greater than $6$. Therefore, seeking a more appropriate DMA weighting matrix is necessary when $N_e$ is large.

Next, we present the estimation accuracy loss raised by the spatial gridding sparsifying dictionary $\mathbf{A}_{\rm G}$. The estimation accuracy is defined as an MSE, i.e., $\varrho= \mathbb{E}\{\| \mathbf{g}^* - \hat{\mathbf{g}} \|_2\}$, where $\mathbf{g}^*$ is the accurate channel. 
When applying the spatial gridding sparsifying dictionary, the sparse channel is obtained by assuming the pilot signals are incident from discretized grids. This discretization leads to inevitable sparse representation error, resulting in a significant decrease in estimation accuracy.

\begin{theorem}
\label{theorem_sp}
Let $\varrho$ represent the estimation accuracy loss in the absence of sparsifying dictionary error, and let $\varrho_{\mathbf{A}_{\rm G}}$ denote the estimation accuracy when utilizing the spatial gridding sparsifying dictionary. Then, the estimation accuracy loss raised by the spatial gridding sparsifying dictionary is given by 
\begin{equation}
\begin{aligned}
        \Delta \varrho =& \varrho - \varrho_{\mathbf{A}_{\rm G}}
    =\sum_{l_p=1}^{L_p} a_{l_p}\left(\text{erf} \left( \frac{\xi - a_{l_p} }{\sqrt{2}\sigma}\right) - \text{erf} \left( \frac{\xi - a_{l_p}\chi_{l_p}}{\sqrt{2}\sigma}\right)\right), 
\end{aligned}
\end{equation}
where $\chi_{l_p} = \mathbf{a}^{\rm H}\left(\hat{\theta}_{g}, \hat{\phi}_{g}\right)\mathbf{a}\left(\theta_{l_p}, \phi_{l_p}\right)$; $\hat{\theta}_{l_p}$ and $\hat{\phi}_{l_p}$ are the azimuth and elevation angle of the gird that is closest the incident direction of $l_p$-th path.
\end{theorem}
\begin{IEEEproof}
Please refer to Appendix \ref{AppendixB}.
\end{IEEEproof}

Theorem \ref{theorem_sp} elucidates that the estimation accuracy loss, $\Delta\varrho$, is contingent upon the spatial gridding interval. A sparser grid leads to a heightened estimation accuracy loss. Conversely, as the grid becomes denser, $\chi_{l_p} \rightarrow 1$ and $\Delta\varrho \rightarrow 0$. However, an increase in grid density also introduces a greater computational burden. In conclusion, the spatial gridding sparsifying dictionary faces a tradeoff between estimation accuracy and computational burden. To break this, an efficient way is to integrate the channel sparse character and learn a sparsifying dictionary from channel data.

From Theorems \ref{thm:unsuccessweight} and \ref{theorem_sp}, we can conclude that it is essential to optimize $\mathbf{Q}$ and $\mathbf{A}$ to achieve higher channel estimation accuracy before solving the problem \eqref{eq_lasso}. 

\subsection{LISTA with Sensing Matrix Optimization}
\label{subsection-LISTA-SMO}

In order to further improve the accuracy of sparse recovery, in this section we propose a novel LISTA-sensing matrix optimization (SMO) algorithm to jointly train the LISTA network and optimize the sensing matrix, i.e., optimizing the DMA weighting matrix $\mathbf{Q}$ and the sparse dictionary $\mathbf{A}$.

To jointly optimize $\mathbf{Q}$ and $\mathbf{A}$, we have the following optimization problem
\begin{equation}
\begin{aligned}
~\min_{\mathbf{Q} , \mathbf{A}}~& { \| \mathbf{g}^*- \mathbf{A}\mathcal{F}(\mathbf{z}|\mathbf{Q} ,\mathbf{A})  \|_2} \\
~\text{s.t} ~& \mathbf{z} =  \mathbf{Q}  \mathbf{H} \mathbf{y},\\
 & \mathbf{y} = \mathbf{g}^* + \mathbf{n},\\
 & q_{n, l}\in \mathbb{Q} =\left\{\frac{j+e^{j \varphi}}{2} \mid \varphi \in[0,2 \pi]\right\}  ,
\end{aligned}
\label{eq_optimizedQA}
\end{equation}
where $\boldsymbol{\alpha} = \mathcal{F}(\mathbf{z}|\mathbf{Q}, \mathbf{A})$ represents the input and output relationship of CS algorithm; $\mathbf{g}^*$ is the actual channel; $\mathbf{n}$ is additive noise. Different from \eqref{LASSO}, the independent variable of this problem is $\mathbf{y}$, as $\mathbf{z}$ is the function of $\QQ$.

Note that it is quite challenging to solve \eqref{eq_optimizedQA} due to the following two reasons. First, $\mathcal{F}(\mathbf{z}|\mathbf{Q},\mathbf{A})$ does not exist an analytical expression. Second, the optimal solution to \eqref{eq_optimizedQA} depends on the instantaneous channel information, which is unknown before carrying out channel estimation. 

To tackle these challenges,  we seek a statistical optimization solution based on collected channel data. Therefore, we reformulate  problem \eqref{eq_optimizedQA} as 
\begin{equation}
\begin{aligned}
\min_{\mathbf{Q} , \mathbf{A}} ~& \sum_{s=1}^{S} \| \mathbf{g}^*_s-\mathbf{A}\mathcal{F}(\mathbf{z}_s|\mathbf{Q} ,\mathbf{A})   \|_2 \\
~\text{s.t} ~& \mathbf{z}_s =  \mathbf{Q}  \mathbf{H} \mathbf{y}_s ,\\
& \mathbf{y}_s = \mathbf{g}^*_s + \mathbf{n},~s=1,2,\cdots,S,\\
 &     q_{n, l}\in \mathbb{Q} =\left\{\frac{j+e^{j \varphi}}{2} \mid \varphi \in[0,2 \pi]\right\}  ,
\end{aligned}
\label{eq_statisticaloptimizationQA}
\end{equation}
where $\mathbf{y}_s$ and $\mathbf{g}^*_s$ are the sample pairs from the collected channel data set.

% Compared with \eqref{eq_optimizedQA}, 

Problem \eqref{eq_statisticaloptimizationQA} enables us to optimize/train the DMA weighting matrix $\mathbf{Q}$  and the sparse
dictionary $\mathbf{A}$ in an offline manner, which is not only easy to implement but also more meaningful from the perspective of compressed sensing.

To solve \eqref{eq_statisticaloptimizationQA}, we draw on the concept of deep unfolding and propose a new model-based algorithm named LISTA-SMO.  LISTA-SMO unfolds the DMA signal processing and sparse representation into trainable layers with parameters symbolizing the DMA weighting matrix and sparsifying dictionary, respectively.
These two layers are referred to as the DMA layer and the sparsifying representation layer. They are integrated into LISTA, facilitating the simultaneous optimization of matrices $\mathbf{Q}$ and $\mathbf{A}$ and training the LISTA recovery network. 

\begin{figure*}[!h]
\centering
\includegraphics[width=0.9\textwidth]{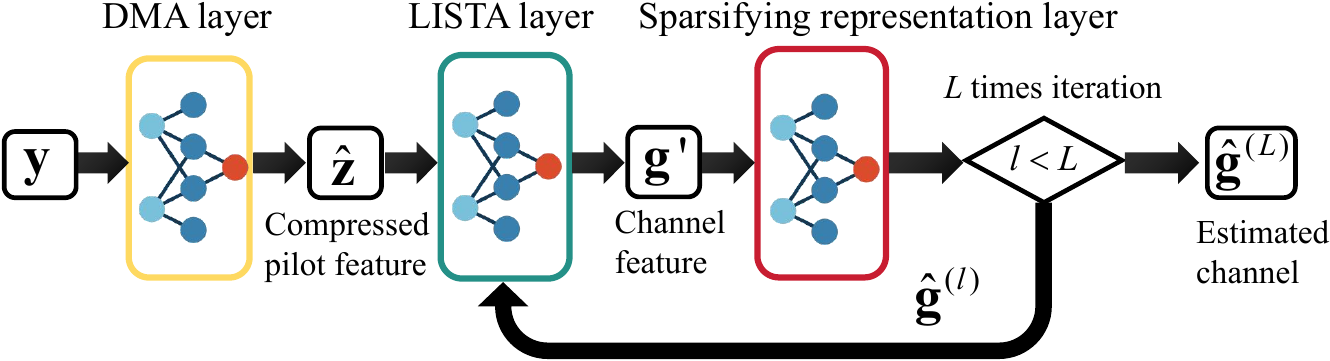}
\caption{The illustration of LISTA-SMO}
\label{fig-block——diagram——LISTA-SMO}
\end{figure*}

Specifically, LISTA-SMO consists of three layers: the DMA layer, the LISTA layer, and the sparse representation layer, as shown in Fig.~\ref{fig-block——diagram——LISTA-SMO}. Firstly, the DMA layer simulates the signal processing process of DMA with the uncompressed pilot signal $\mathbf{y}$ as the input and the features $\hat{\mathbf{z}}$ as the output. The features $\hat{\mathbf{z}}$ correspond to the received pilot at the end of the RF chain $\mathbf{z}$, which is named the compressed pilot feature. This process merges $\QQ$ into the neural network. Secondly, the LISTA layer utilizes the compressed pilot feature $\hat{\mathbf{z}}$ and the estimated channel $\hat{\mathbf{g}}^{(l-1)}$ of the last iteration to update the channel feature $\mathbf{g}'$. Lastly, the sparsifying representation layer performs the sparse nonlinear function on $\mathbf{g}'$ to generate the $l$-th layer's estimated channel $\hat{\mathbf{g}}^{(l)}$. Next, we will delve into the intricacies of the DMA layer and the sparse representation layer. Then, we propose LISTA-SMO by combining the DMA layer and the sparse represent layer with the LISTA.

\subsubsection{DMA Layer}
\label{subsection-LISTA-WMO}
To optimize the DMA weighting matrix, we unfold the signal processing process of DMA into the DMA layer, where the DMA weighting matrix is constructed as learnable parameters. 

It can be observed from \eqref{Q_constrain} that the configurable parameter of the DMA weighing matrix on each element is $q_{n_d, n_e}$, which depends on its own phase $\varphi_{n_d, n_e}$. Correspondingly, we construct $\mathbf{W}_{\boldsymbol{\varphi}} \in \mathbb{R}^{N \times 1}$ to collect the phases of all antenna elements. Then, the learning parameter ${\mathbf{W}_\mathbf{Q}} \in \mathbb{C}^{N_d \times N}$ (corresponding to $\QQ$) can be constructed as 
\begin{equation}
[{\mathbf{W}_\mathbf{Q}}]_{n_{1},\left(n_{2}-1\right) N_d+n_e} = \left\{\begin{array}{ll}
\left(j+\mathrm{exp}({j \mathcal{S}([{\mathbf{W}_{\varphi}}]_{\left(n_{1}-1\right) N_d+n_e}})\right)/2 & n_{1}=n_{2} \\
0 & n_{1} \neq n_{2}
\end{array} \right.,
\end{equation}
where $\mathcal{S}(x) = \frac{2\pi}{1+e^{-x}}$ is the non-linear function for mapping the definition domain of $\mathbf{W}_{\boldsymbol{\varphi}}$ from $\mathbb{R}$ to $[0, 2\pi]$ (due to $\varphi_{n_d, n_e} \in [0, 2\pi]$). 

The DMA layer formulates the same process as in \eqref{DMA_transmitter}. Correspondingly, the mathematical expression of the DMA layer is  
\begin{equation}
    \hat{\mathbf{z}} = \mathbf{W}_\mathbf{Q}\mathbf{H}\mathbf{y}.
    \label{eq_DMAlayer}
\end{equation}

\subsubsection{Sparsifying Representation Layer}
\label{subsection-LISTA-SDO}
To optimize the sparsifying dictionary, we first construct a learnable sparsifying dictionary $\mathbf{W}_\mathbf{A} \in \mathbb{C}^{N\times D}$. Then, we unfold the sparse representation process into two layers. The one utilizes $\mathbf{W}_\mathbf{A}$ to transform the updated channel feature $\mathbf{g}'$ into the hidden layer feature $\hat{\boldsymbol{\alpha}}$ (corresponding to $\BA$), i.e., $\hat{\BA} = h_{(\hat{\eta})}(\mathbf{W}_\mathbf{A}\mathbf{g}')$. The other utilizes $\mathbf{W}_\mathbf{A}^{{\rm{H}}}$ to recover $\hat{\mathbf{g}}$ with $\hat{\boldsymbol{\alpha}}$, i.e., $\hat{\mathbf{g}} = \mathbf{W}_\mathbf{A}^{{\rm{H}}} \hat{\BA}$. In summary, we can build the sparse representation layer as 
\begin{equation}
    \hat{\CH}^{(l)}=\mathbf{W}_\mathbf{A}^{{\rm{H}}} \cdot h_{(\hat{\eta})}(\mathbf{W}_\mathbf{A}\CH').
    \label{eq-SRL}
\end{equation}

\subsubsection{LISTA-SMO}

% In this part, we detail the LISTA-SMO by integrating the DMA layer and the sparse represent layer into the LISTA.
% Before going to LISTA-SMO, we introduce  of LISTA-SMO, which is
In LISTA-SMO, we apply Analytic Weight LISTA (ALISTA) as the backbone, which is a new form of LISTA \cite{chen_theoretical_2018, osti_10191388}. The layer of ALISTA is given by 
\begin{equation}
\label{ALISTA}
\BA{}^{(l+1)}=h_{(\hat{\eta}^{(l)})}\left(\BA{}^{(l)} - \kappa^{(l)}\mathbf{W}(\mathbf{Q}\mathbf{H}\mathbf{A}\BA{}^{(l)}-\mathbf{z} ) \right),
\end{equation}
where $\mathbf{W}\in \mathbb{C}^{D\times N_d}$, $\hat{\eta}$ and $\kappa$ are learnable parameters.
% and $\mathbf{W}$ is same for all layer. Compared with LISTA, ALISTA is more efficient because it reuses the parameter $\mathbf{W}$ in each layer. 

\begin{figure*}[!h]
\centering
\includegraphics[width=\textwidth]{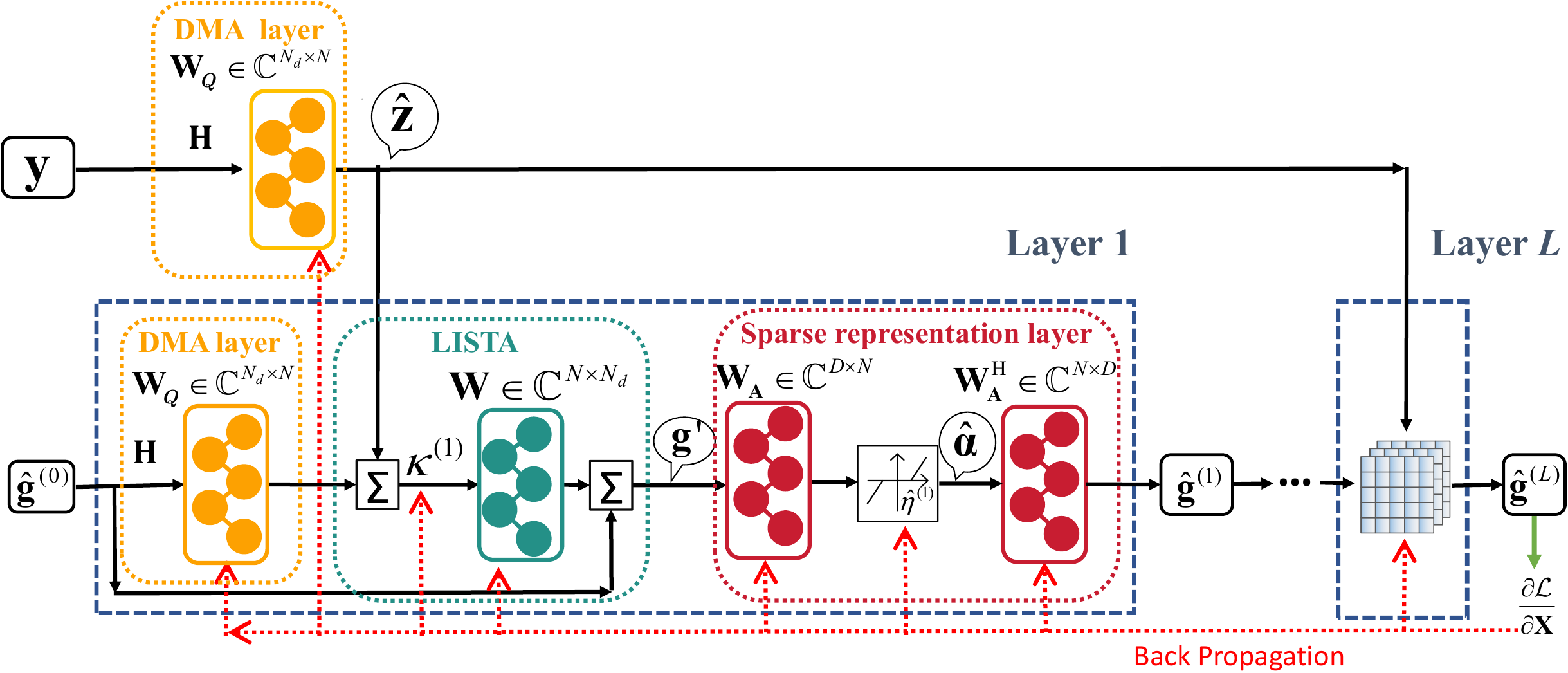}
\caption{The block diagram of LISTA-SMO}
\label{fig-LISTA-SMO}
\end{figure*}

We integrate the DMA layer and the sparse represent layer into the ALISTA and propose the LISTA-SMO, which is shown in Fig.~\ref{fig-LISTA-SMO}. Mathematically, the architecture of LISTA-SMO can be represented as
\begin{equation}
\label{eq_LISTA-SMO}
\hat{\CH}^{(l+1)}=(\mathbf{W}_\mathbf{A})^{{\rm{H}}} \cdot h_{(\hat{\eta}^{(l)})}\left(\mathbf{W}_\mathbf{A}\left(
\hat{\CH}^{(l)} - \kappa^{(l)}\mathbf{W}_{\rm SMO}(\mathbf{W}_\mathbf{Q}\mathbf{H}\hat{\CH}^{(l)}-\hat{\mathbf{z}}\right)\right).
\end{equation}

By comparing \eqref{ALISTA} and \eqref{eq_LISTA-SMO}, the integration contains three aspects. On the first aspect, LISTA-SMO modifies two computational operations to integrate the DMA layer into ALISTA. The first modification replaces $\mathbf{z}$ with $\hat{\mathbf{z}} = \mathbf{W}_\mathbf{Q}\mathbf{H}\mathbf{y}$ for simulating the DMA process, which is also the DMA layer in Fig.~\ref{fig-block——diagram——LISTA-SMO}. This DMA layer is the upper DMA layer in Figure \ref{fig-LISTA-SMO}. Besides, to improve the training efficiency, we replace $\mathbf{Q}$ in \eqref{ALISTA} with $\mathbf{W}_\mathbf{Q}$ that embeds the DMA layer inside each layer, which is shown with the yellow part inside the LISTA-SMO layer. On the second aspect, to optimize the sparsifying dictionary, LISTA-SMO incorporates \eqref{eq-SRL} by introducing $\mathbf{W}_\mathbf{A}$ and $\mathbf{W}_\mathbf{A}^{{\rm{H}}}$ before and after the soft shrinkage function, which is shown in red part of Fig.~\ref{fig-LISTA-SMO}. The LISTA layer is shown in the green box. On the third aspect, different from the ALISTA that updates $\boldsymbol{\alpha}$, the LISTA layer in LISTA-SMO directly updates the channel with the operation $\CH' =\mathbf{W}_\mathbf{Q}\mathbf{H}\CH^{(l)}-\hat{\mathbf{z}}$. Compared with LISTA, the updated feature in LISTA-SMO changes from $\boldsymbol{\alpha}$ to $\mathbf{g}$, which demands the recovery parameter $\mathbf{W}$ of LISTA change to $\mathbf{W}_{\rm SMO}= \mathbf{W}_\mathbf{A}^{-1} \mathbf{W}$. 

The LISTA-SMO contains the forward and backward processes, which are represented with the dark line and red dashed lines in Fig.~\ref{fig-LISTA-SMO}, respectively. The forward process feeds ${\bf{y}}$ to the neural network and directly outputs the estimated channel $\mathbf{g}^{(L)}$. The backward process updates the parameters by minimizing the loss function on the training data set. The training data set is $\mathbb{D}_{\text{SMO}} = \left\{[\mathbf{y}_s, \mathbf{g}^*_s]| \mathbf{y}_s = \mathbf{g}^*_s + \mathbf{n}_{\rm{m}}, s = 1,...,S \right\}$, $\mathbf{y}_s$ and $\mathbf{g}^*_s$ are the sample pair of $\mathbb{D}_{\text{SMO}}$. $ \mathbf{n}_{\rm{m}}$ is the measurement noise. The loss function is 
\begin{equation}
    \mathcal{L} = \sum_{s=1}^S\| \mathbf{g}^*_s - \mathbf{g}^{(L)}_s \|_2,
    \label{eq_loss_function_smo}
\end{equation}
where $\mathbf{g}^{(L)}_s$ is the output corresponding to $\mathbf{y}_s$.

During the training stage, LISTA-SMO will update the DMA weighting matrix, the sparsifying dictionary, and the LISTA parameters. After that, we can configure the optimized $\mathbf{W}_\mathbf{Q}$ on the DMA. Meanwhile, we remove the first DMA layer and directly send $\mathbf{z}$ into each layer to estimate the channel. We conclude the proposed LISTA-SMO in Algorithm \ref{alg3}.

\begin{algorithm}[h]
\caption{LISTA-SMO for DMA channel estimation. }
\label{alg3}
\begin{algorithmic}[1]
\STATE \textbf{Training Process}
\STATE \textbf{Initialize:} Initialize the parameter $\mathbf{W}$, $\mathbf{W}_\mathbf{Q}$, $\mathbf{W}_\mathbf{A}$, $\kappa$ and $\hat{\eta}$ with random number. 
\STATE \textbf{Input:} The data set $\mathbb{D}_{\text{SMO}} = \left\{[\mathbf{y}_s, \mathbf{g}^*_s]| \mathbf{y}_s = \mathbf{g}^*_s + \mathbf{n}, s = 1,...,S \right\}$. 
\STATE \textbf{While} 
\STATE \quad Sample a batch of data from the data set $\mathbb{D}_{\text{SMO}}$;
\STATE \quad Calculate the output of neural network $\hat{\CH}^{(L)}$ and calculate the loss $\mathcal{L}$ in (\ref{eq_loss_function_smo});
\STATE \quad Update the parameter $\mathbf{W}$, $\mathbf{W}_\mathbf{Q}$, $\mathbf{W}_\mathbf{A}$, $\kappa$ and $\hat{\eta}$. 
\STATE \textbf{Until} $\mathcal{L}$ converge. 
\STATE \textbf{Inference Process}
\STATE Configure the optimized $\mathbf{W}_\mathbf{\varphi}$ on the DMA and remove the first DMA layer.
\STATE Feed the receive pilot $\mathbf{z}$ into the well-trained neural network $\mathcal{N}$ to estimate the channel $\mathbf{g}' = \mathbf{g}^{(L)}$.
\end{algorithmic}
\end{algorithm}

\begin{remark}
\label{remark-WMO}
The DMA layer is used to optimize $\mathbf{Q}$ in order to enhance the signal originating from specific angles with high incident probabilities. 
\end{remark}

% Comment \ref{remark-WMO} illustrates the feasibility of optimizing the DMA weighting matrix.

\begin{remark}
\label{remark-LISTA-SDO}
The sparse representation layer is capable of achieving higher channel estimation accuracy than LISTA because the loss function for LISTA-SMO is directly calculated by the channel, which avoids introducing the sparse representation error at the beginning.
\end{remark}

\begin{remark}
\label{remark-computation_efficience}
LISTA and LISTA-SMO both share the advantage of requiring a relatively small number of parameters. Specifically, the number of parameters for LISTA is $2L+ D^2 + N_dD$, while for LISTA-SMO it is $2L+ DN+ 2N_dN$. This indicates that LISTA-SMO only adds an additional $N_dN$ parameter compared to LISTA when $D=N$.
% While this results in an increase in parameters, the simulation results presented in Section IV demonstrate that both LISTA-SMO and LISTA maintain similar convergence times, suggesting that the extra parameters do not significantly impact computational efficiency.
\end{remark}

\subsection{Self-supervised Learning for LISTA-SMO}
\label{subsection-self-supervisedlearning}

Constructing the date set $\mathbb{D}_{\text{SMO}}$ may be challenging due to the complexity of obtaining noiseless signal $\mathbf{g}^*$ from wireless communication systems. In this subsection, we propose a self-supervised learning method to address this challenge. Self-supervised learning represents an innovative learning approach that trains a neural network similar to supervised learning but with a key distinction. Rather than relying on manually annotated supervised labels, self-supervised learning automatically generates its labels based on the inherent characteristics of the training samples.

By capitalizing on the physical significance of the training data, we develop the self-supervised learning version of LISTA-SMO. Specifically, from  \eqref{eq_loss_function_smo}, it can be seen that the estimation precision of our algorithm is contingent upon the accuracy of the label $\mathbf{g}^*$. We assume $\mathbf{g}^*$ in $\mathbb{D}_{\text{SMO}}$ are noised measurements, the lower boundary of the average estimation accuracy $\varrho_{\rm{SMO}}$ for our algorithm is 
\begin{equation}
    \begin{aligned}
        \varrho_{\rm{SMO}} &= \| (\mathbf{g}^*_s + \mathbf{n}_{\rm{m}}) - \mathbf{g}^{(L)}_s \|_2 \geq \sigma^2_{\rm{m}},
    \end{aligned}
    \label{lowerboundary}
\end{equation}
where $ \mathbf{n}_{\rm{m}}$ is the measurement noise and $\sigma^2_{\rm{m}}$ is the variance of noise. 

Equation \eqref{lowerboundary} reveals that the performance of our algorithm is lower-bounded by the accuracy of the channel measurements. \eqref{lowerboundary} also suggests that LISTA-SMO can be trained with imperfect channel data. Given the fact that the input of LISTA-SMO is the channel measurement, i.e., $\mathbf{y} = \mathbf{g}^* + \mathbf{n}_{\rm{m}}$, we can directly substitute it with $\mathbf{y}_s$ and train the neural network with the following objective function:
\begin{equation}
\mathcal{L}= \sum_{s=1}^{S} \| \mathbf{y}_s -\mathbf{g}^{(L)}_s \|_2.
\label{loss5}
\end{equation} 

Equation \eqref{loss5} indicates that training the neural network is a self-supervised method as it leverages the inputs as labels instead of accurate labels.  

\begin{remark}
The self-supervised learning method only needs the measurement channel data $\mathbf{y}_s$, rather than the noiseless data $\mathbf{g}^*$, which directly eases the pressure on the data collection, making the process more efficient.
\end{remark}

\section{Numerical Results}
\label{sec:result}
In this section, we provide numerical results to verify the channel estimation accuracy of our proposed algorithms. 
We present the simulation setup and the data set in Subsection \ref{subsection-simulation_setup}. Then, we numerically evaluate the performance of our algorithm in \ref{performance_comparison}. 
% Next, we compare the NMSE of our algorithm with different numbers of layers in \ref{subsection-differentlayer}. Besides, we compare the NMSE performance between our algorithms and competing alternatives in Subsection \ref{subsection-SparsifyingDictionary}.
% . Finally, Subsection \ref{subsection-CompressedRatio} show the NMSE of our algorithms in different compressed ratios.

\subsection{Simulation Setup}
\label{subsection-simulation_setup}

\subsubsection{DMA Parameter Setup}
We consider a planar DMA shown in Fig.~\ref{systemmodel} throughout the experiment. The DMA comprises $N_d = 20$ microstrip. Each microstrip accommodates $N_e = 20$ elements. The DMA works with the carrier frequency 28 GHz ($\lambda$ =  1.07 cm). The element spaces are $\Delta d_x = \lambda/2$ and $\Delta d_y = \lambda/2$. We use $\alpha = 0.6 \text{~m}^{-1}$ and $\beta=827.67 \mathrm{~m}^{-1} $ to represent the propagation inside the DMA waveguides, assuming a microstrip implemented in Duroid 5880 with 30 mill thickness \cite{wang_dynamic_2021}. 

\subsubsection{Channel Data Generation }

The channel $\mathbf{g}$ is randomly generated following the channel model detailed in \eqref{channel_model}. In that model, the number of channel paths $L_p$ is randomly chosen from 2 to 6. The channel path gain is generated by  
\begin{equation}
a_l\left(d\right)=\sqrt{F\left(\phi_{l_p}\right)} \frac{\lambda}{4\pi d},
\end{equation}
where $d$ is the path length generated by a uniform distribution $U(15, 60)$. $F\left(\phi_{l_p}\right)$ is the radiation profile of each element, defined as  $F\left(\phi_{l_p}\right)= 2(b+1) \cos ^{b}\left(\phi_{l_p}\right)$, with $b=2$ denoting the Boresight gain. The incident angles $\theta_{l_p}$ and $\phi_{l_p}$ are randomly sampled from the uniform distribution, which are
$ \left[(\phi, \theta)|\phi \sim \mathcal{U}(\frac{\pi}{12}, \frac{5\pi}{12}), \theta \sim \mathcal{U}(\frac{\pi}{6}, \frac{5\pi}{6})\right]$.
The sparse representation of channel $\boldsymbol{\alpha}$ is obtained by assuming the channel is incident from the nearest grid.

We build the training, test, and validation sets with the randomly generated channel data. The training, test, and validation sets include 512000, 512, and 51200 channel data, respectively. The training set and test set are utilized in the training stage, whose sample has random SNR to increase its robustness to noise. The validation set is for testing the NMSE when the SNR is $[0, 3, \ldots, 21]$ dB.

\subsubsection{Algorithm Parameter Setup}
We quantify the performance of algorithms by the normalized MSE (NMSE) on the  validation set, given by
\begin{equation}
    \mathrm{NMSE}=\mathbb{E}\left\{\frac{\left\|\mathbf{g}^*-\mathbf{g}^{(L)}\right\|_{2}^{2} }{\|\mathbf{g}^*\|_{2}^{2}}\right\}.
\end{equation}

Unless otherwise stated, the LISTA and LISTA-SMO leverage the following parameters.
In the aspect of the sparsifying dictionary, the LISTA leverages a space-grids sparsifying dictionary, in which the coverage area is uniformly partitioned into $D = 400$ girds, in which the $\varphi$ axis is partitioned into 20 parts and $\theta$ axis is partitioned into 20 parts. Mathematically, the grid can be formulated as
\begin{equation}
    [(\varphi, \theta)|\varphi = k_\varphi\frac{\pi}{40}, \theta = k_\theta\frac{\pi}{20}, 1 \leq k_\varphi \leq 20, 1 \leq k_\theta \leq 20,  k_\varphi\in \mathbb{Z}, k_\theta\in \mathbb{Z}].
    \label{eq_gridding}
\end{equation}
Correspondingly, the LISTA-SMO applies $\mathbf{W}_{\mathbf{A}}$ with 400 atoms. Both LISTA and LISTA-SMO contain 16 layers and utilize the Adam optimizer with the same learning rate as 1e-4. The initial value of $\mathbf{W}_a, \mathbf{W}_b, \mathbf{W}_{\BPHI}, \mathbf{W}_{\mathbf{A}}, \mathbf{W}$ are random generated. The initial value of $\hat{\eta}$ and $\kappa$ are $10^{-4 }$ and $1$, respectively.

\subsection{Performance Comparison}
\label{performance_comparison}
\begin{figure*}[!h]
\centering
\subfloat[LISTA\label{result-lista}]{\includegraphics[width=0.4\textwidth]{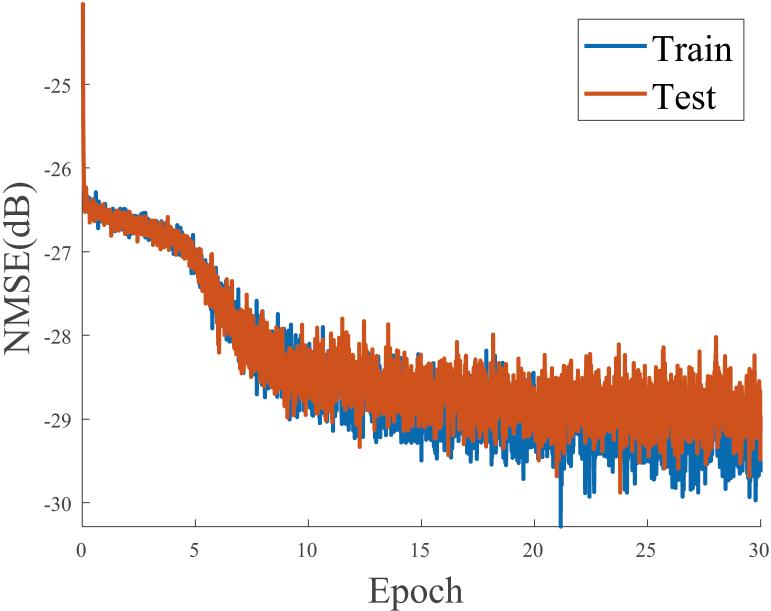} }
% \hfill
\subfloat[ISTA-NET]{\includegraphics[width=0.4\textwidth]{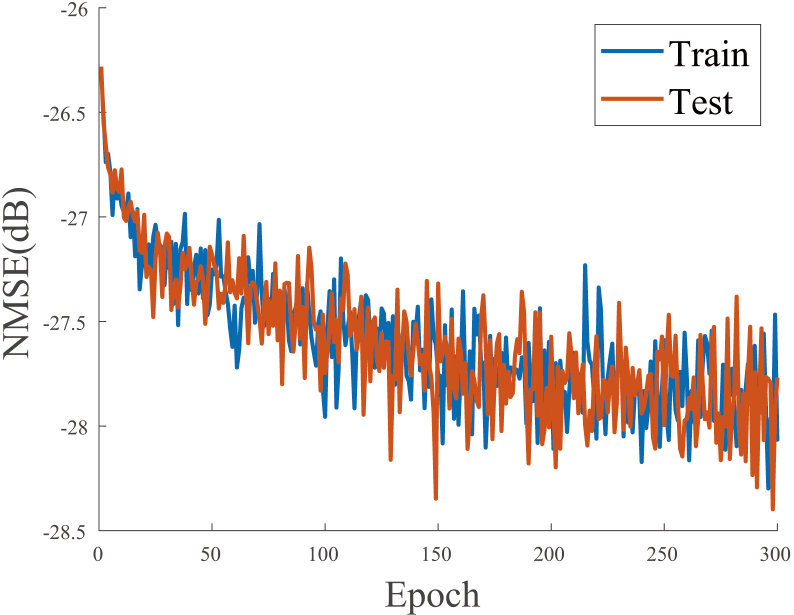}}\\
\subfloat[LISTA-SMO(supervised)\label{result-dce}]{\includegraphics[width=0.4\textwidth]{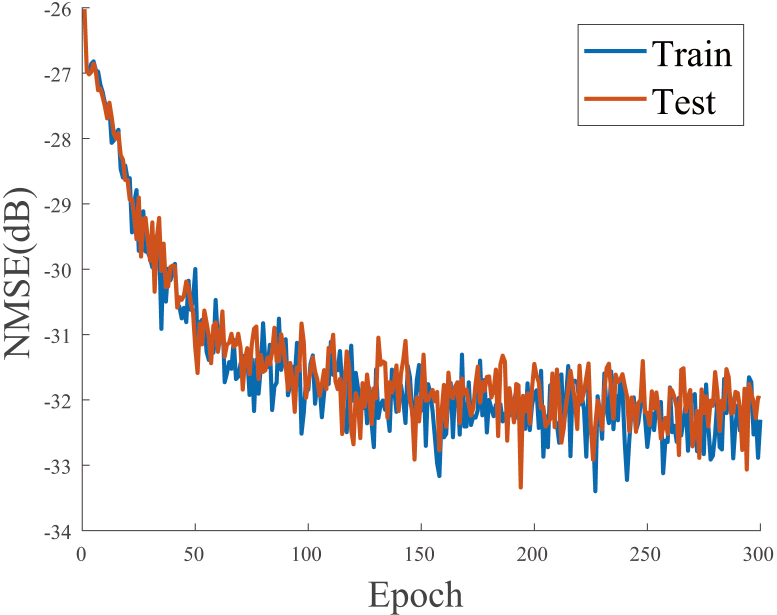}}
% \hfill
\subfloat[LISTA-SMO(self-supervised learning)\label{result-self_supervised}]{\includegraphics[width=0.4\textwidth]{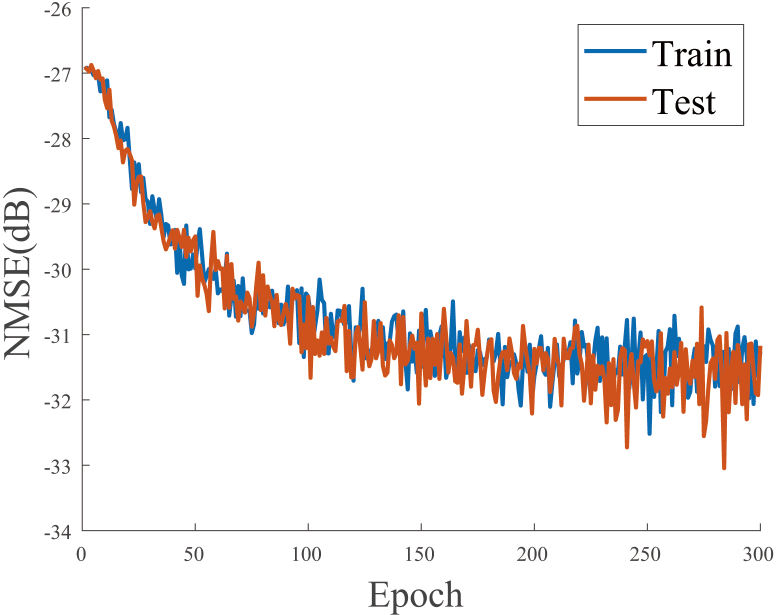}}
\caption{The convergence curve of different algorithms. (a) LISTA. (b) ISTA-NET. (c) LISTA-SMO (supervised). (d). LISTA-SMO (self-supervised learning).}
\label{convergence_curve}
\end{figure*}

First, we show the efficiency of model-based learning algorithms by comparing the convergence performance of LISTA, ISTA-NET, and LISTA-SMO in Fig.~\ref{convergence_curve}. During the training stage, we sample data from the training set without repetition and define an epoch as all the samples being sampled once. We define convergence as the point where the mean of NMSE of the test set in the last ten iterative times is no longer descending. In Fig.~\ref{convergence_curve}, the blue and red lines represent the NMSE on the training and test sets, respectively.

\label{sec_coverageperformance}

Fig.~\ref{convergence_curve} illustrates the superior convergence performance of both LISTA and LISTA-SMO compared to ISTA-NET in terms of convergence speed. Specifically, LISTA demonstrates exceptional efficiency, achieving convergence in approximately ten epochs in Fig.~\ref{convergence_curve}(a). In contrast, LISTA-SMO, as shown in Fig.~\ref{convergence_curve}(c), requires around 100 epochs to converge due to the optimization of two matrices. As shown in Fig.~\ref{convergence_curve}(b), ISTA-NET, which demands more than 200 epochs to achieve convergence. By comparing Fig.~\ref{convergence_curve}(b) and (C), LISTA-SMO outperforms ISTA-NET in both training efficiency and estimation accuracy. Furthermore, we investigate the convergence behavior of LISTA-SMO in the self-supervised learning scenario, as shown in Fig.~\ref{convergence_curve}\subref{result-self_supervised}. Remarkably, LISTA-SMO (self-supervised) exhibits a similar convergence speed to its supervised counterpart while maintaining comparable estimation accuracy. These results suggest that our proposed self-supervised learning method remains effective even without utilizing explicit labels.

\label{subsection-differentlayer}
\begin{figure}[!h]
\centering
\includegraphics[width=0.5\textwidth]{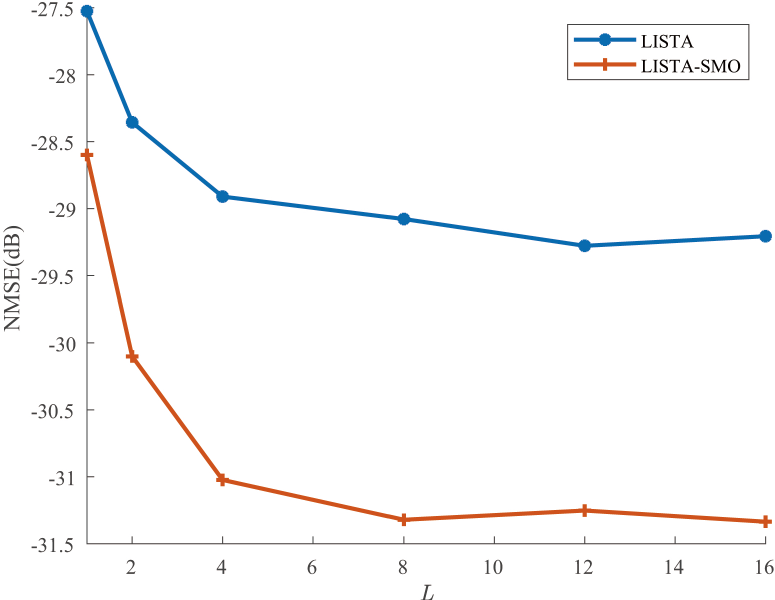}
\caption{The NMSE versus the number of layers $L$}
\label{layercompare}
\end{figure}

Next, we study the impact of the number of layers on the NMSE achieved by LISTA and LISTA-SMO.  In Fig.~\ref{layercompare}, the $x$ axis represents for different numbers of layers, i.e., $L=[1, 2, 4, 8, 12, 16]$ and the $y$ axis represents the mean NMSE on the test set. From Fig.~\ref{layercompare}, it is observed that both LISTA and LISTA-SMO exhibit a decreasing trend in NMSE as the number of layers increases. However, an interesting observation is that both LISTA and LISTA-SMO require a minimum of eight layers to achieve optimal performance. This indicates that LISTA-SMO keeps the same training complexity as LISTA, through the LISTA-SMO having more parameters.

\begin{figure}[!hb]
\centering
\includegraphics[width=0.5\textwidth]{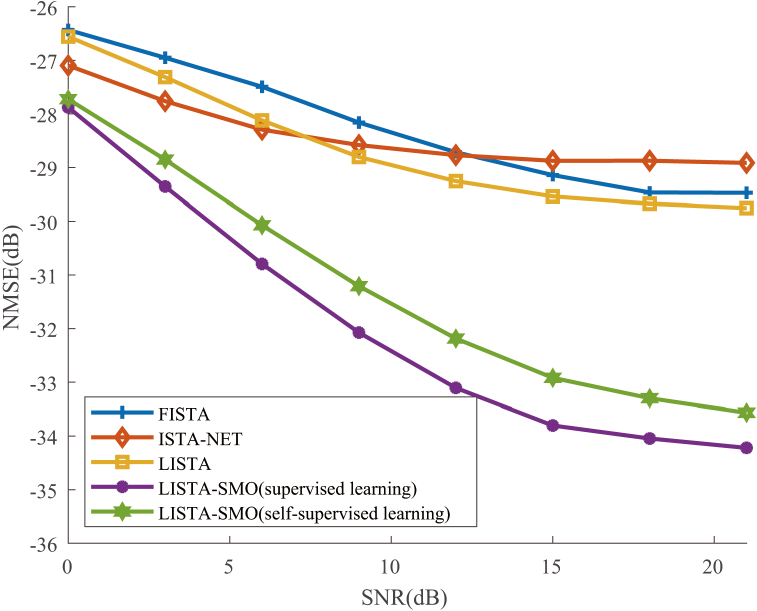}
\caption{The NMSE versus SNR.}
\label{algorithmcompare}
\end{figure}

In Fig.~\ref{algorithmcompare}, we compare the NMSE performance of our proposed algorithms with that of two benchmark algorithms: FISTA and ISTA-Net, under different values of SNR. The FISTA algorithm is the new version of ISTA, which has faster convergence \cite{beck2009fast}. In our experiments, both FISTA and ISTA-Net utilize the random DMA matrix and the spatial gridding sparsifying dictionary. From Fig.~\ref{algorithmcompare}, it is observed that as the SNR increases, the NMSE of all algorithms decreases as expected. However, it is clearly observed that our proposed LISTA-SMO with supervised learning has significant superiority in overall algorithms, e.g., with an almost $6$ dB improvement over FISTA, $5$ dB improvement over LISTA, and $4$ dB improvement over ISTA-NET, indicating that the optimization of the DMA weighting matrix and the sparsifying dictionary achieves higher estimation accuracy. Moreover, we can also observe from Fig.~\ref{algorithmcompare} that the NMSE of LISTA-SMO with self-supervised learning is slightly higher than that of LISTA-SMO with supervised learning. This is because the LISTA-SMO cannot recover some weak channel components without the help of labels.

\begin{figure}[!hb]
\centering
\includegraphics[width=0.5\textwidth]{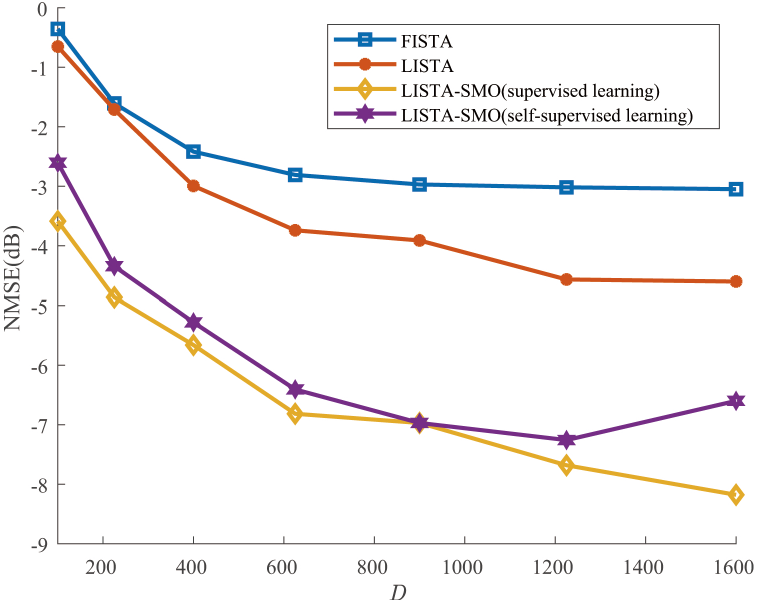}
\caption{The NMSE versus the sparsifying dictionary size.}
\label{Sdcompare}
\end{figure}

\begin{figure}[!hb]
\centering
\includegraphics[width=0.5\textwidth]{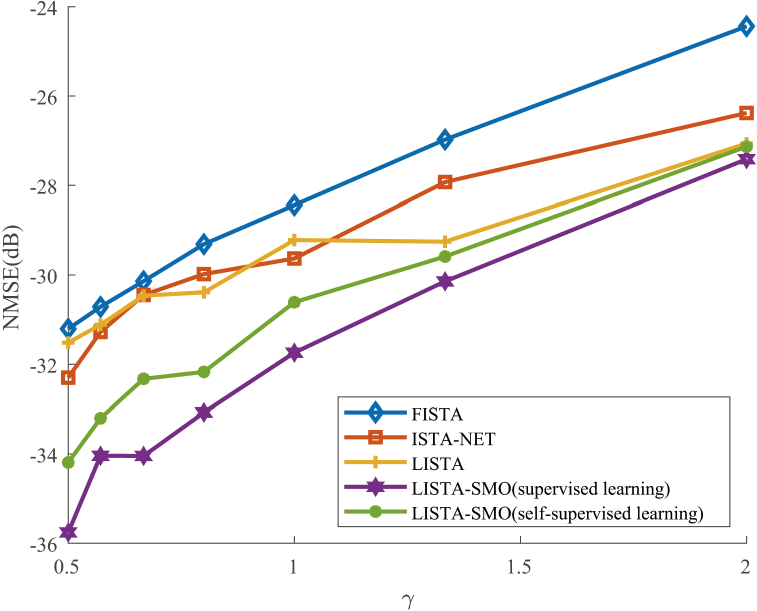}
\caption{The NMSE versus the compressed ratio $\gamma$.}
\label{result-DMAsetting}
\end{figure}

In Fig.~\ref{Sdcompare}, we study the impact of the sparsifying dictionary size on the achieved NMSE by different algorithms. In our experiments, the grids of the spatial gridding dictionary are obtained by uniformly partitioning the $\theta$ axis and $\varphi$ axis into $10$, $15$, $20$, $25$, $30$, $35$, and $40$ parts. Thus, the dictionary sizes are $D=[100, 225, 400, 625, 900, 1225, 1600]$. From Fig.~\ref{Sdcompare}, it is evident that the NMSE achieved by all algorithms generally decreases as the dictionary size $D$ is small and eventually saturates as $D$ becomes large. For instance, when $D=100$, all algorithms underutilize the available degrees of freedom, leading to suboptimal performance. In such cases, increasing $D$ significantly reduces the NMSE. However, when $D>900$, all algorithms reach their peak performance, and further increases in $D$ do not enhance channel estimation accuracy. Moreover, Fig.~\ref{Sdcompare} demonstrates that LISTA-SMO outperforms other algorithms consistently. For example, LISTA-SMO with $D=400$ achieves nearly $3$ dB improvement over ISTA-NET and LISTA with $D=1600$. Furthermore, we observe a slightly higher NMSE for LISTA-SMO with self-supervised learning compared to LISTA-SMO with supervised learning.
% , indicating that the self-supervised approach may face challenges in recovering weak channel components without labeled data.

Finally, we study the impact of the compressed ratio on the NMSE. The compressed ratio is defined as $\gamma = \frac{N_e}{P N_d}$, where $P$ is the number of pilots. Fig.~\ref{result-DMAsetting} shows the NMSE achieved by our proposed and benchmark algorithms when $\gamma = [0.68, 0.76, 0.86, 1, 1.17, 1.42, 1.81, 2.5]$. From Fig.~\ref{result-DMAsetting},  it is clearly observed that as the compressed ratio $\gamma$ increases, the NMSE of all algorithms displays an upward trend. This finding underscores the significance of the compressed ratio in relation to the NMSE. In real-world systems, if the number of RF chains is intentionally limited to minimize hardware costs and power consumption, a  valuable approach involves effectively increasing the number of pilots to recover the channel with the necessary accuracy.

\section{Conclusion}
\label{sec:conclusion}
This paper studied the channel estimation problem in DMA-assisted MIMO systems. We first formulated it as a CS problem by sparsifying the channel with the space-gridding sparsifying dictionary. We solved the CS problem with an efficient model-based algorithm, LISTA, under the random DMA weighting matrix. However, the DMA weighting matrix and sparsifying dictionary could result in decreased channel estimation accuracy, so we further proposed a model-based learning algorithm, LISTA-SMO, to optimize the DMA weighting matrix and SD. LISTA-SMO is the LISTA neural network appended with the DMA layer and the sparse representation layer, which are built by unfolding the DMA signal process and the sparsifying dictionary into neural network layers. The DMA layer and the sparse representation layer contain the DMA weighting matrix and sparsifying dictionary as 
learnable parameters, and thereby, the optimized matrices can be optimized while LISTA training. To reduce the requirement of the data set, we also proposed a self-supervised learning method to train LISTA-SMO. Simulation results demonstrated that the model-based algorithm, LISTA, outperforms the other channel estimation algorithms, and LISTA-SMO significantly overperformed LISTA. Our findings highlight the importance of optimizing the DMA weighting matrix and sparsifying dictionary and the effectiveness of our algorithms.

\appendices
\section{Proof of Theorem \ref{thm:unsuccessweight}}
\label{AppendixA}
\begin{IEEEproof}
% Theorem \ref{thm:unsuccessweight} is mainly for demonstrating the influence of the DMA weighting matrix.
In order to prove Theorem \ref{thm:unsuccessweight}, we first introduce some definitions of CS. 
\begin{definition}[Definition 1.3 of \cite{eldar2012compressed}]
\label{RIP}
The sensing matrix $\mathbf{\Psi}\in\mathbb{C}^{N_d \times D}$ satisfies the Restricted Isometry Property (RIP) of order $k$ if there exists a $\delta_{k} \in(0,1)$ such that 
\begin{equation}
\delta_k(\mathbf{\Psi})=\max _{\mathscr{T} \subset \mathscr{N}, \operatorname{card}(\mathscr{T}) \leq k}\left\|\mathbf{\Psi}_\mathscr{T}^{\mathrm{H}} \mathbf{\Psi}_\mathscr{T}-\mathbf{I}\right\|_{2}
\label{RIC}
\end{equation}
where $\mathscr{N}$ represents the set $\{1,\cdots, N\}$; $\mathscr{T}$ is the subset of $\mathscr{N}$; 
$\operatorname{card}(\mathscr{T})$ is the cardinality of $\mathscr{T}$. $\mathbf{\Psi}_\mathscr{T}\in\mathbb{C}^{N_d \times \operatorname{card}(\mathscr{T})}$ is the submatrix of $\mathbf{\Psi}$ that is constituted by the column vectors of $\mathbf{\Psi}$ with indexes in $\mathscr{T}$; $\mathbf{I}$ is the unit matrix.
\end{definition}
\begin{lemma}[Theorem 1.9 of \cite{eldar2012compressed}]
\label{extra_recover}
Supporting $\|\boldsymbol{\alpha}\|_0 \leq k$ and the sensing matrix $\mathbf{\Psi}$ satisfies the RIP of order $2k$ with $\delta_{2 k}<\sqrt{2}-1$, $\boldsymbol{\alpha}$ can be accurately recovered from $\mathbf{z}$. 
\end{lemma}

Based on Lemma \ref{extra_recover}, we can conclude that, to recover a channel that contains $2L_p$ channel paths component, $\mathbf{\Psi} = \mathbf{QHA}$ must satisfy the $2L_p$ order RIP, i.e., $\delta_{2L_p}<\sqrt{2}-1$. Hence, in this proof, we aim to provide the probability of $\delta_{2L_p}<\sqrt{2}-1$ when each non-zero element of the random DMA weight matrix $\mathbf{Q}_{\rm Ran}$ have the random phase $\varphi \sim \mathcal{U}[0, 2\pi]$.

Because we concentrate solely on the impact of the DMA weighting matrix, we employ a unitary matrix as a sparse dictionary to simplify the analysis, i.e., $\mathbf{A}^{\rm{H}}\mathbf{A} = \mathbf{I}$. Then, we have 
\begin{equation}
\begin{aligned}
        \left\|\mathbf{\Psi}_\mathscr{T}^{\rm{H}}\mathbf{\Psi}_\mathscr{T} - \mathbf{I}\right\|_2 &= \left\|\mathbf{A}_\mathscr{T}^{\rm{H}}(\mathbf{Q}_{\rm Ran}\mathbf{H})^{\rm{H}}(\mathbf{Q}_{\rm Ran}\mathbf{H})\mathbf{A}_\mathscr{T} - \mathbf{I}\right\|_2 \\
         & = \left\|\mathbf{A}_\mathscr{T}^{\rm{H}}\mathbf{C}\mathbf{A}_\mathscr{T} - \mathbf{I}\right\|_2 \\
         & = \left\|\mathbf{C}_\mathscr{T} - \mathbf{I}\right\|_2 
        % & = \left\|\mathbf{C}_\mathscr{T}^{\rm{H}}\mathbf{C}_\mathscr{T} - \mathbf{I}\right\|_2. 
\end{aligned}
\label{eq-appendix-psi2q}
\end{equation}
where $\mathbf{A}_\mathscr{T}$ as the submatrix of $\mathbf{A}$ that is constituted by the column vectors of $\mathbf{A}$ with index $\mathscr{T}$; $\mathbf{C} = (\mathbf{Q}_{\rm Ran}\mathbf{H})^{\rm{H}}(\mathbf{Q}_{\rm Ran}\mathbf{H})$; $\mathbf{C}_\mathscr{T} = \mathbf{A}_\mathscr{T}^{\rm{H}}(\mathbf{Q}_{\rm Ran}\mathbf{H})^{\rm{H}}(\mathbf{Q}_{\rm Ran}\mathbf{H})\mathbf{A}_\mathscr{T}$. By substituting \eqref{eq-appendix-psi2q} to \eqref{RIC}, we have
\begin{equation}
    \delta_{2L_p} = \max _{\mathscr{T} \subset \mathscr{N}, \operatorname{card}(\mathscr{T}) \leq 2L_p}\left\|\mathbf{C}_\mathscr{T} - \mathbf{I}\right\|_{2}.
    \label{eq-appendix-RICC}
\end{equation}
\eqref{eq-appendix-RICC} illustrates that $\delta_{2L_p}$ is the maximum eigenvalue of $\mathbf{C}_\mathscr{T} - \mathbf{I}$, where $\operatorname{card}(\mathscr{T}) = 2L_p$. 

Given that the columns of $\mathbf{A}_\mathscr{T}$ are orthogonal, the eigenvalue set of $\mathbf{C}$ is the subset of the eigenvalue set of $\mathbf{C}_\mathscr{T}$. Mathematically, it can be represented as $\mathscr{S}_{\mathbf{C}_\mathscr{T}} \subset \mathscr{S}_{\mathbf{C}}$, $\operatorname{card}(\mathscr{S}_{\mathbf{C}_\mathscr{T}}) = 2L_p$, in which $\mathscr{S}_{\mathbf{C}}$ and $\mathscr{S}_{\mathbf{C}_\mathscr{T}}$ as the eigenvalue set of $\mathbf{C}$ and $\mathbf{C}_\mathscr{T}$, respectively. By examining the eigenvalues set $\mathscr{S}_{\mathbf{C}}$, we can subsequently discern the eigenvalue properties of matrix $\mathbf{C}_\mathscr{T}$. 

Due to the specific architecture of DMA in $\eqref{Q_constrain}$, we have 
\begin{equation}
    \mathscr{S}_\mathbf{C} = \left\{\sum_{l=1}^{N_e}\|q_{1,l}h_{1,l}\|_2, \cdots ,\sum_{l=1}^{N_e}\|q_{N_d,l}h_{N_d,l}\|_2\right\}.
\end{equation}

When the phase of element's weighting value $\varphi_{n,l}$ follows $\mathcal{U}[0, 2\pi]$, $\|q_{n,l}h_{n,l}\|_2$ follows uniform distribution $\mathcal{U}[0, 1]$. The $n$-th elements of $\mathscr{S}(\mathbf{C})$ can be approximated as $s_n = \sum_{l=1}^{N_e}\|q_{n,l}h_{n,l}\|_2 \sim \mathcal{N}(N_e/2, N_e/12)$, as $N_e \gg 1$ (according to the central limit theorem). Thereby, $\mathscr{S}(\mathbf{C})$ is a set of independent and identically distributed random variables whose elements have the cumulative distribution function $\mathcal{F}(s_n) = \frac{1}{2}\left[1 + \text{erf} \left( \frac{s_n - N_e/2}{\sqrt{2}N_e/12} \right)\right]$, in which $\text{erf}$ is the error function. 

Such as this, we represent the maximum value of $\mathscr{S}_{\mathbf{C}_\mathscr{T}}$ as $s_{\max}$. Since $\mathscr{S}_{\mathbf{C}_\mathscr{T}}$ is constituted by random $2L_p$ elements in $\mathscr{S}_\mathbf{C}$, the cumulative distribution function of $s_{\max}$ is $\mathcal{F}({s_{\max}}) = (\mathcal{F}(s_n)) ^ {2L_p}$. Then, the probability of $s_{\max}<\sqrt{2}-1$ is $\left[\frac{1}{2} + \frac{1}{2}\text{erf} \left( \frac{\sqrt{2} - N_e/2}{\sqrt{2}N_e/12} \right)\right]^{2L_p}$. As $\delta_{2L_p}$ is $s_{\max}$, the probability that the sparse vector $\mathbf{\alpha}$ can be \textit{exactly} recovered from the noiseless pilot with a probability $p_{\rm rec} = \left[\frac{1}{2} + \frac{1}{2}\text{erf} \left( \frac{\sqrt{2} - N_e/2}{\sqrt{2}N_e/12} \right)\right]^{2L_p}$, which complete the proof.
% To sum up, when the phase of $q$ follows the phase of element's weighting value $\varphi_{n,l}$ follows $U[0, 2\pi]$, the algorithm can recover the sparse signal $\boldsymbol{\alpha}$ from the received pilot $\mathbf{z}$ with the probability of $\left[\frac{1}{2} + \frac{1}{2}\text{erf} \left( \frac{\sqrt{2} - N_e/2}{\sqrt{2}N_e/12} \right)\right]^{2L_p}$. Theorem \ref{thm:unsuccessweight} is proofed. 

\end{IEEEproof}

\section{Proof of Theorem \ref{theorem_sp}}
\label{AppendixB}
\begin{IEEEproof}
In this proof, we discuss the impact of the spatial gridding sparsifying
dictionary to the estimation accuracy loss $\varrho= \mathbb{E}\{\| \mathbf{g}^* - \hat{\mathbf{g}} \|_2\}$. 

We assume the CS algorithm perfectly solves of \eqref{eq_lasso} and the solution $\hat{\boldsymbol{\alpha}}$  following the Karush-Kuhn-Tucker (KKT) condition. According to the subgradient objective of the KKT condition, we have 
\begin{equation}
 \begin{array}{ll}
\mathbf{\Psi}^{\rm H}(\mathbf{z}-\mathbf{\Psi} \hat{\boldsymbol{\alpha}})=\xi \boldsymbol{\beta},
\label{eq-solution}
 \end{array}
\end{equation}
where $\xi$ is the regularization parameter in \eqref{eq_lasso}; $\boldsymbol{\beta} = [\boldsymbol{\beta}_1, \cdots, \boldsymbol{\beta}_D]$, in which $\boldsymbol{\beta}_i = 1$ when $\hat{\boldsymbol{\alpha}} \neq 0$
and $\boldsymbol{\beta}_i < 1$ when $\hat{\boldsymbol{\alpha}}=0$.

As this proof only focuses on the impact of $\mathbf{A}_{\rm G}$, we ignore the DMA weighting matrix and rewrite \eqref{eq-solution} as 
\begin{equation}
    \mathbf{A}_{\rm G}^{\rm H}(\mathbf{y}-\mathbf{A}_{\rm G} \hat{\boldsymbol{\alpha}})=\xi \boldsymbol{\beta}.
    \label{eq-appendix-q-lasso}
\end{equation}
From the literature \cite{osborne2000lasso}, we have the solution of \eqref{eq-appendix-q-lasso}, which is 
\begin{equation}
\begin{aligned}
& \hat{\boldsymbol{\alpha}}_{\backslash \mathscr{M}}=0, \\
& \hat{\boldsymbol{\alpha}}_{\mathscr{M}}=\left(\mathbf{A}_{\mathscr{M}}^{\rm H} \mathbf{A}_{\mathscr{M}}\right)^{-1}\left[\mathbf{A}_{\mathscr{M}}^{\rm H} \mathbf{y}-\xi\mathbf{I}\right],
\label{eq-appendix-solution-lasso}
\end{aligned}
\end{equation}
in which $\mathscr{M}$ is the set that contains the indexes of $\mathbf{a}_i^{\rm H}  \mathbf{y}> \xi$, where $\mathbf{a}_i$ is the $i$-th atom of $\mathbf{A}_{\rm G}$; $\boldsymbol{\alpha}_\mathscr{M}$ is vector that is constituted by the non-zero element of $\boldsymbol{\alpha}$; $\mathbf{A}_\mathscr{M}$ is the submatrix of $\mathbf{A}_{\rm G}$ that the columns $\mathbf{A}_\mathscr{M}$ is constituted by $\mathbf{A}_{\rm G}$ with index $\mathscr{M}$. 

Equation \eqref{eq-appendix-solution-lasso} illustrates that the lasso solution shrinks $\mathbf{A}_{\rm G}^{\rm H}\mathbf{y}$. The solution is shrunk to zero, when $\mathbf{a}_i^{\rm H}  \mathbf{y}< \xi$. In our channel estimation problem, the channel can be represented by several channel path components, i.e., $\mathbf{y} = \mathbf{g}^* + \mathbf{n} = \sum_{l_p=1}^{L_p} a_{l_p} \mathbf{a}\left(\theta_{l_p}, \phi_{l_p}\right)+ \mathbf{n}$. A channel path component can be kept as long as there is one atom to let $\mathbf{a}_i^{\rm H}\mathbf{y}>\xi$. In contrast, the channel path component will be shrunk to zero when $\max \left(\mathscr{E}  \right)  <\xi$, where 
$$
\mathscr{E} = \left\{a_{l_p} \mathbf{a}_1^{\rm H}\mathbf{a}\left(\theta_{l_p}, \phi_{l_p}\right) + \mathbf{a}_i^{\rm H}\mathbf{n}, \cdots, a_{l_p} \mathbf{a}_D^{\rm H}\mathbf{a}\left(\theta_{l_p}, \phi_{l_p}\right) + \mathbf{a}_D^{\rm H}\mathbf{n}\right\}. 
$$
We denote $\mathscr{E}_m$ as the maximum element of $\mathscr{E}$. $\mathscr{E}_m$ will be shown at the channel path component and its closest atom. 

We define $\chi = \mathbf{a}^{\rm H}\left(\hat{\theta}_{i}, \hat{\phi}_{i}\right)\mathbf{a}\left(\tilde{\theta}_{l_p}, \tilde{\phi}_{l_p}\right)$ as the mismatched error, in which $\hat{\theta}_{l_p}$ and $\hat{\phi}_{l_p}$ are the azimuth and elevation angle of the atom that is closest the incident direction of $l_p$-th path.
Then, as $\mathbf{n}\sim\mathcal{N}(0,\sigma^2)$, the probability that $l_p$-th channel path component can not be recovered, i.e., $\mathscr{E}_m<\xi$, is $\mathcal{F}_{l_p}(\xi) = \frac{1}{2}\left[1 + \text{erf} \left( \frac{\xi - a_{l_p}\chi }{\sqrt{2}\sigma} \right)\right]$, in which $\text{erf}$ is the error function. When the $l_p$-th channel path component is shrunk, the estimation accuracy loss $\varrho$ will raise $a_{l_p}^2$. By counting the $L_p$ path, the estimation accuracy loss $\varrho$ follows normal distribution with the expectation is $\mathbb{E}\{\varrho\} = \sum_{l_p=1}^{L_p} a_{l_p} \mathcal{F}_{l_p}(\xi)$ and the variance is $\operatorname{Var}\{ \varrho\} = \sum_{l_p=1}^{L_p} \left[a_{l_p}^2 \mathcal{F}_{l_p}(\xi) - a_{l_p}^2 \mathcal{F}_{l_p}^2(\xi) \right]$ (according to the central limit theorem). 

To show the impact of the sparsifying dictionary, we show the difference in the probability distribution
of estimation accuracy loss $\varrho$ when $\chi$ is different. It is clear that there is no error brought by the sparsifying dictionary when $\chi = 1$. Thereby, when $\chi <1$, the difference of the expectation of the estimation accuracy loss is 
\begin{equation}
    \Delta \varrho = \sum_{l_p=1}^{L_p} a_{l_p}\left(\text{erf} \left( \frac{\xi - a_{l_p} }{\sqrt{2}\sigma}\right) - \text{erf} \left( \frac{\xi - a_{l_p}\chi_{l_p}}{\sqrt{2}\sigma}\right)\right). 
\end{equation}

Hence, the proof is completed.
\end{IEEEproof}

\bibliographystyle{IEEEtran}
\bibliography{ref}

\end{document}